%% file: access.tex
\documentclass{IEEEtran}

\usepackage{hyperref}

\usepackage{lipsum}

\usepackage{booktabs}
\usepackage{cite}
\usepackage{amsmath,amssymb,amsfonts}
\usepackage{algorithmic}
\usepackage{graphicx}
\usepackage{textcomp}
\usepackage{makecell}
\usepackage{url}
\usepackage{adjustbox}
\usepackage{caption}
\usepackage{subcaption}
\usepackage{soul}
\def\BibTeX{{\rm B\kern-.05em{\sc i\kern-.025em b}\kern-.08em
    T\kern-.1667em\lower.7ex\hbox{E}\kern-.125emX}}

\usepackage[none]{hyphenat}

\begin{document}

\title{A Comprehensive Comparative Study of Individual ML Models and Ensemble Strategies for Network Intrusion Detection Systems}

\author{Ismail Bibers, Osvaldo Arreche, and Mustafa Abdallah 
\thanks{This work is partially supported by the Lilly Endowment through the AnalytixIN grant, the Enhanced Mentoring Program with Opportunities for Ways to Excel in Research (EMPOWER), and the 1st Year Research Immersion Program (1RIP) grants from the Office of the Vice Chancellor for Research at Indiana University-Purdue University Indianapolis. Ismail Bibers and Mustafa Abdallah are with Computer and Information Technology Department, Purdue University in Indianapolis, Indianapolis, IN, USA. Email: {\tt \{ibibers,abdalla0\}@purdue.edu}. Osvaldo Arreche is with Electrical and Computer Engineering Department, Purdue University in Indianapolis, Indianapolis, IN, USA. Email: \tt {oarreche@purdue.edu}}}  



\maketitle
\IEEEpeerreviewmaketitle

\thispagestyle{empty}
\pagestyle{empty}

\begin{abstract}
The escalating frequency of intrusions in networked systems has spurred the exploration of new research avenues in devising artificial intelligence (AI) techniques for intrusion detection systems (IDS). Various AI techniques have been used to automate network intrusion detection tasks, yet each model possesses distinct strengths and weaknesses. Selecting the optimal model for a given dataset can pose a challenge, necessitating the exploration of ensemble methods to enhance generalization and applicability in network intrusion detection. This paper addresses this gap by conducting a comprehensive evaluation of diverse individual models and both simple and advanced ensemble methods for network IDS. We introduce an ensemble learning framework tailored for assessing individual models and ensemble methods in network intrusion detection tasks. Our framework encompasses the loading of input datasets, training of individual models and ensemble methods, and the generation of evaluation metrics. Furthermore, we incorporate all features across individual models and ensemble techniques. The study presents results for our framework, encompassing 14 methods, including various bagging, stacking, blending, and boosting techniques applied to multiple base learners such as decision trees, neural networks, and among others. We evaluate the framework using two distinct network intrusion datasets, RoEduNet-SIMARGL2021 and CICIDS-2017, each possessing unique characteristics. Additionally, we categorize AI models based on their performances on our evaluation metrics and via their confusion matrices. Our assessment demonstrates the efficacy of learning across most setups explored in this study. Furthermore, we contribute to the community by releasing our source codes, providing a foundational ensemble learning framework for network intrusion detection.
\end{abstract}



\sloppy


\input{Introduction}

\input{Related_work}
\input{Background}
\input{Framework}

\input{Evaluation}
\input{Discussion}

\input{Conclusion}

\bibliographystyle{IEEEtran}
\bibliography{refs}

\appendices
\input{Appendix}

\end{document}

%% file: Introduction.tex
\section{Introduction}

The primary aim of intrusion detection systems (IDS) is to detect unauthorized utilization, misuse, and exploitation of computer network systems by both internal users and external intruders~\cite{northcutt2002network,mukherjee1994network,apruzzese2022modeling}. Traditional IDS designs typically operate under the assumption that the behavior of an intruder will deviate noticeably from that of a legitimate user and that many unauthorized actions are discernible. The potential of artificial intelligence (AI) has spurred the advancement of fully automated intrusion detection systems~\cite{buczak2015survey,dina2021intrusion}. Various AI methodologies have been employed to automate intrusion detection tasks, including neural networks~\cite{kim2017method,tang2020saae}, decision trees~\cite{ferrag2020rdtids,al2021intelligent}, logistic regression~\cite{nick2007logistic,panigrahi2022intrusion}, and random forest~\cite{balyan2022hybrid,waskle2020intrusion}.

The majority of these AI methods, with the exception of random forest, operate as standalone learning models where the combination of their decisions is not utilized by the IDS~\cite{arisdakessian2022survey,sabev2020integrated}. Each of these AI models harbors its own constraints, such as a high false positive rate for certain models (for instance, approximately half of the major companies contend with 10,000 security alerts daily from AI-based threat monitoring tools~\cite{AI_False_Rates}), and a high false negative rate for others (which poses a significant challenge in safety-critical computer network applications~\cite{mijalkovic2022reducing}). 

Prior AI-focused studies predominantly emphasized the classification accuracy of different AI algorithms without harnessing the collective potential of these diverse AI techniques. This inherent limitation has stressed the urgent necessity to exploit various ensemble learning methods to bolster IDS~\cite{al2021classification,aburomman2017survey,tama2021ensemble}.

Numerous recent studies have begun delving into the utilization of ensemble learning with various AI models for IDS, as evidenced by works such as~\cite{LAZZARINI2023110941,future-ensemble,novel-ensemble-nids,kitsune,Class-and-clus-et,ensemble-libraries,EC-NIDS,NIDS-EML,NIDS-Analysis-EL-FS,Multi-dimensional-ensemble-NIDS,Twd-On-NIDS-EL,Ensemble-Learning-Framework-IDS-IOT,Ensemble-IDS-MLA,NIDS-Emsemble-CVM-FS}. Specifically, works like~\cite{novel-ensemble-nids,NIDS-EML,NIDS-Analysis-EL-FS,Twd-On-NIDS-EL,Ensemble-Learning-Framework-IDS-IOT,kitsune,ensemble-libraries} have proposed ensemble learning frameworks for anomaly detection, focusing on binary classification to discern normal from anomalous traffic. Conversely, other studies~\cite{LAZZARINI2023110941,future-ensemble,EC-NIDS,Ensemble-IDS-MLA,Multi-dimensional-ensemble-NIDS,NIDS-Emsemble-CVM-FS} have developed ensemble learning frameworks for classification of network intrusions, encompassing categories like Denial of Service (DoS) attacks, Port Scanning, Normal traffic, and others.

These frameworks employ ensemble learning techniques such as Boosting, Stacking, and Bagging, considering various base models like Decision Trees, Support Vector Machines, and Neural Networks. Primary evaluation metrics include traditional AI metrics like accuracy, precision, recall (true positive rate), F1 score, and false positive rates. While most studies utilize benchmark datasets for IDS, such as CICIDS-2017, KDD'99, NSL-KDD, and UNSW-NB15~\cite{novel-ensemble-nids,NIDS-EML,NIDS-Analysis-EL-FS,Ensemble-IDS-MLA}, some conduct tests on real networks like the Palo Alto network~\cite{Twd-On-NIDS-EL}, and even in real-time scenarios, as demonstrated by the ``kitsune'' framework~\cite{kitsune}.

An exemplary contribution in this domain is the work by~\cite{future-ensemble}, which focuses on generating a new dataset and benchmarking it using ensemble learning techniques. Another notable approach is showcased in~\cite{ensemble-libraries}, where ensemble procedures are employed to select the AI model variant with the best performance. However, a comprehensive evaluation of a wide array of AI methods across different intrusion datasets is lacking in these works, potentially impacting their general applicability. Each study tends to concentrate on a singular ensemble learning method to enhance the performance of a limited set of base models.

This paper aims to address the aforementioned gap by comprehensively evaluating diverse ensemble methods for network intrusion detection systems. We establish multiple Individual ML models and Simple and Advanced ensemble learning frameworks to assess such methods in the context of network intrusion detection. Leveraging prior works such as~\cite{LAZZARINI2023110941,future-ensemble,novel-ensemble-nids,kitsune,Class-and-clus-et,ensemble-libraries,EC-NIDS,NIDS-EML,NIDS-Analysis-EL-FS,Multi-dimensional-ensemble-NIDS,Twd-On-NIDS-EL,Ensemble-Learning-Framework-IDS-IOT,Ensemble-IDS-MLA,NIDS-Emsemble-CVM-FS}, which have outlined various ensemble learning approaches, our framework can be categorized as follows. 

\begin{itemize}
\item \textbf{Individual Models}: The initial phase of the framework involves implementing individual models such as decision trees~\cite{ferrag2020rdtids,al2021intelligent}, logistic regression~\cite{nick2007logistic,panigrahi2022intrusion}, and neural networks~\cite{kim2017method,tang2020saae}. This phase encompasses tasks like loading datasets (e.g., CICIDS-2017, and RoEduNet-SIMARGL2021), training the models, and assessing performance using metrics like accuracy, precision, recall, and F1 score. 

\item \textbf{Simple Ensemble Methods}: The subsequent stage of the framework involves implementing simple ensemble methods such as averaging, max voting, and weighted averaging. Performance evaluation is conducted using metrics like accuracy, precision, recall, and F1 score.

\item \textbf{Advanced Ensemble Methods}: The third phase focuses on implementing advanced ensemble methods including bagging, boosting, stacking, and blending. Note that we consider random forest~\cite{balyan2022hybrid,waskle2020intrusion} in this category since it is built based on bagging of many decision trees.  Again, evaluation metrics like accuracy, precision, recall, and F1 score are used to assess performance.

\item \textbf{Comparative Analysis}: The final step entails evaluating all individual, simple, and advanced ensemble models to identify the most effective models for IDS and analyze the impacts of ensemble learning techniques.
\end{itemize}


Additionally, our study presents results for various ensemble model combinations, including bagging methods, stacking, and boosting, applied to multiple base learners such as decision trees, logistic regression, random forest, neural networks, among others. These distinctions highlight the novel contributions of our work compared to previous studies, as discussed in Related Work Section (Section~\ref{sec:related_work}).

We conduct evaluations of our framework using two prominent network intrusion datasets, each with distinct characteristics. The first dataset, RoEduNet-SIMARGL2021~\cite{mihailescu2021proposition}, is a recent collection from the SIMARGL project, supported by the European Union. Notably, to our knowledge, limited prior works has applied comprehensive ensemble learning methods to this dataset, as discussed in the related work section. This dataset comprises realistic network traffic data, including features derived from live traffic, rendering it highly suitable for network intrusion detection systems. The second dataset utilized in our evaluation is CICIDS-2017~\cite{panigrahi2018detailed}, established by the Canadian Institute for Cybersecurity at the University of Brunswick in 2017. This dataset serves as a benchmark for intrusion detection and encompasses various attack profiles.

For each dataset, we assess various approaches, encompassing different base learners and variants of ensemble methods, applied to different AI models. The AI models under consideration include Logistic Regression (LR), Decision Trees (DT), K-Nearest Neighbors (KNN), Multi-Layer Perceptron (MLP), Adaptive Boosting (ADA), eXtreme Gradient Boosting (XGB), CatBoosting (CAT), Gradient Boosting (GB), Averaging (Avg), Max Voting, Weighted Averaging, and Random Forest (RF).

For all these models across both datasets, we present and analyze the evaluation metrics generated by our framework. We thoroughly discuss the results, providing insights into the performance of each method. Additionally, we categorize the AI models based on their performances on evaluation metrics with the datasets considered in this study. Notably, we rank these different methods in descending order of F1 score, offering a clear perspective on their effectiveness (given by performance on the network intrusion datasets).

This comprehensive evaluation allows us to identify the most promising approaches for network intrusion detection across different datasets and AI models, facilitating informed decision-making in the implementation of IDS. This work represents a significant advancement in bridging the gap in the application of ensemble learning methods for network intrusion detection systems (IDS). Through conducting extensive evaluations and comparisons of various metrics, we contribute to enhancing the understanding of these ensemble methods' efficacy in the realm of IDS.

The metrics employed in our evaluation encompass crucial network security requirements for AI models, including accuracy, precision, recall, and F1 score of intrusion detection methods, along with their corresponding runtimes. By thoroughly examining these metrics, we provide valuable insights into the performance and efficiency of different ensemble methods in detecting network intrusions. Our framework not only addresses existing limitations but also expands the application of ensemble learning techniques in network intrusion detection systems. By doing so, we pave the way for further advancements and enhancements in this research area, ultimately contributing to the development of more robust and effective network security solutions. 

\noindent \textbf{Summary of Contributions:}
We summarize below our main contributions in this current work.\vspace{-1pt} 
\begin{itemize}
    \item \textbf{Evaluation of Individual and Ensemble Learning Methods:} We conduct a comprehensive evaluation and comparison of various Individual ML models, along with various simple and advanced ensemble learning methods for network intrusion detection systems (IDS).
    
    \item \textbf{Assessment Across Diverse Metrics:} Our evaluation considers a range of metrics crucial for network security requirements, including accuracy, precision, recall, and F1 score of intrusion detection methods, as well as their runtime performance.
        
    \item \textbf{Evaluation on Two Prominent Datasets:} We evaluate our framework on two well-known network intrusion datasets with distinct characteristics: RoEduNet-SIMARGL2021 and CICIDS-2017. This allowed for a comprehensive analysis across different network intrusion scenarios.
    
    \item \textbf{Performance Ranking:} We categorized AI models (individual and ensemble ones) based on their performances on evaluation metrics, ranking these methods in descending order of F1 score, providing valuable insights into the effectiveness of each approach.
    
    \item \textbf{Expansion of Ensemble Learning Applications for IDS:} By demonstrating the efficacy of ensemble learning methods in network IDS, our work expands the application of these techniques in this critical research area, paving the way for further advancements.
    
    \item \textbf{Availability of Source Codes:} We make our source codes available to the community for accessing the framework designed for network intrusion detection and for further development with new datasets and models.\footnote{The GitHub URL for our  source codes is: 
\url{https://github.com/sm3a96/IDS-Machine-Learning-Techniques-.git}}

\end{itemize}

%% file: Related_work.tex
\vspace{-1mm}
\section{Related Work}\label{sec:related_work}

\subsection{Existing Efforts in Leveraging Ensemble Learning for IDS}
The survey conducted in the previous work~\cite{Class-and-clus-et} offers an overview of intrusion detection systems (IDS), focusing on the evolution of ensemble systems and methodologies employed in their design, particularly emphasizing ensemble techniques between 2009 and 2020. This study comprehensively discusses the current state of ensemble models, highlighting various approaches such as Stacking, Bagging, Boosting, and voting, among others. The analyzed works encompass a range of datasets, including KDD'99, NSL-KDD, Kyoto 2006+, and AWID, along with diverse models such as neural networks (NN), support vector machines (SVM), and decision trees (DT), fuzzy clustering, and radial basis function (RBF). The primary contribution of this work lies in its in-depth exploration of the existing landscape, stimulating the investigation of novel combination methods, such as the exploration of new combination rules. These insights offer valuable directions for further research on ensemble learning for IDS.

\textbf{Ensemble Learning for Binary Classification Anomaly Detection Approaches:} Within this domain, the study by \cite{novel-ensemble-nids} introduces an anomaly detection framework that operates on input datasets such as CICIDS-2017, UNSW-NB15, and KDD'99. The framework preprocesses the data and conducts feature selection (employing the Chi-square method in this study), subsequently applying various base models including Gaussian Naive Bayes, Logistic Regression, and Decision Trees. The predictions are then integrated using the Stochastic Gradient Descent ensemble model to yield the final prediction. The primary contributions of this research lie in the amalgamation of learning algorithms via stacking to enhance IDS performance, with potential applicability to other benchmark datasets. However, the study acknowledges limitations related to data imbalance issues, suggesting that the utilization of data augmentation techniques could alleviate such imbalances. Furthermore, the framework could benefit from incorporating different ensemble learning models to further enhance performance.

Similarly, \cite{NIDS-EML} proposes an ensemble learning framework for binary classification of anomalies in IDS, utilizing the NSL-KDD and UNSW-NB15 datasets along with base models like Random Forest, AdaBoost, XGBoost, and Gradient boosting decision trees. The framework combines the outcomes of these models using a soft voting scheme. The presented results highlight the potential of the proposed NIDS framework to improve the accuracy of cyber-attack detection and minimize false alarm rates.

Moreover, \cite{NIDS-Analysis-EL-FS} presents a framework for IDS applied to datasets such as NSL-KDD, and UNSW-NB15. Base models encompass LR, DT, NB, NN, and SVM, while ensemble techniques include Majority Voting, DT, NB, LR, NN, and SVM. The study also explores combinations of feature selection methods, with results indicating superior overall performance for ensemble techniques. However, the authors underscore the need for new datasets, particularly real-world ones, and advocate for the integration of unsupervised learning methods.

Additionally, \cite{Twd-On-NIDS-EL} applies its framework to real-world datasets, including the Palo Alto network log, in addition to NSL-KDD and UNSW-NB15 datasets. Anomaly detection is addressed through ensemble methods employing weighted voting atop base learners like SVM, Autoencoder, and Random Forest. The primary contribution of this work is the introduction of a new ANIDS approach with real-world applicability, reducing false predictions. However, scalability issues and reliance solely on weighted voting are acknowledged as limitations, potentially necessitating more diverse approaches for efficient performance across different scenarios in this network intrusion detection task.

Considering the IoT domain, the study by \cite{Ensemble-Learning-Framework-IDS-IOT} introduces a framework for anomaly detection utilizing the TON-IoT network dataset. It employs four supervised machine learning (ML) models as base models, including Random Forests, Decision Trees, Logistic Regression, and K-Nearest Neighbors. These base models are subsequently integrated into an ensemble method, employing stacking and voting mechanisms to enhance attack detection efficiency. Limitations of this research include its narrow focus on the TON-IoT dataset without exploring other datasets, and its omission of other popular ensemble learning methods like bagging and averaging.

In contrast, \cite{kitsune} presents an online anomaly detection system for network intrusion detection using a series of ensemble Autoencoders, catering to real-time detection requirements. This approach differs from previous works by leveraging ensemble techniques specifically tailored for Autoencoders. Additionally, \cite{ensemble-libraries} addresses the issue of overfitting in ensemble learning for small binary classification datasets. However, the framework utilizes several non-IDS-related datasets and employs base models such as Random Forest, Naive Bayes, and Logistic Regressor. The primary contribution of this work lies in its ensemble model selection procedure, which searches for the best model for a particular instance. Nonetheless, a limitation of this approach is the high computational cost associated with the cross-validation technique. While pruning may mitigate overfitting, its effectiveness may vary across different datasets and models.

\textbf{Ensemble Learning for Multiclass Classification Approaches:} Several studies explore ensemble learning techniques for multiclass classification in IDS. For example, in~\cite{LAZZARINI2023110941}, a novel approach applies ensemble learning on datasets like CICIDS-2017 and ToN\_IoT. They utilize stacking with Tensorflow models (CNN, DNN, RNN, LSTM), where class predictions feed into a DNN ensemble method. However, limitations include the absence of real IoT scenario experimentation, reliance on a single ensemble method, and resource-intensive operations on IoT devices. Another work,~\cite{future-ensemble}, introduces the GTCS dataset for multiclass classification, addressing NSL-KDD dataset limitations. Employing the Weka toolkit, it employs adaptive ensemble learning with J48, MLP, and IBK base models, utilizing majority voting for ensemble learning. Drawbacks include lack of real-world deployment, external dataset validation, and limited AI model variety.

In~\cite{EC-NIDS}, an ensemble learning approach achieves higher accuracy and lower false alarms by incorporating Random Forest to alleviate data imbalance. Utilizing Linear Genetic Programming (LGP), Adaptive Neural Fuzzy Inference System (ANFIS), and Random Forest classifiers, it employs weighted voting for ensemble. However, challenges include lack of optimal weight assignment, generalization issues across datasets, and model variety. Similarly,~\cite{Ensemble-IDS-MLA} applies bagging to NB, PART, and Adaptive Boosting on KDD'99, using voting for component selection. Limitations involve restricted dataset testing and AI model diversity.

\subsection{Contribution of Our Work}
Our contribution lies in introducing a comprehensive intrusion classification framework encompassing individual ML models, along with simple and advanced ensemble techniques. We operate on two distinct datasets, RoEduNet-SIMARGL2021 and CICIDS-2017, aiming to generate key performance metrics including Accuracy, Recall, Precision, and F1 score. Upon dataset loading and preparation, we incorporate all available features for analysis. Initially, our framework executes individual models leveraging LR, DT, RF, MLP, and KNN as base learners. Subsequently, we explore simple ensemble techniques such as Averaging (Avg), Max Voting, and Weighted Averaging. In the next phase, advanced ensemble methods including Bagging, Boosting methods (ADA, GB, XGB, CAT), Blending, and Stacking are applied in our framework.

Throughout the experimentation, we meticulously collect and evaluate result metrics to benchmark optimal performance. Notably, our work stands out for its extensive benchmarking experimentation across various model combinations. Furthermore, our inclusion of RoEduNet-SIMARGL2021 dataset in the experiments fills a gap in existing research, as prior works seldom consider this dataset in their analyses. 

%% file: Background.tex
\section{Background and Problem Statement}\label{sec:background}

This section outlines the fundamental concepts of network intrusion detection, highlights the hurdles posed by artificial intelligence (AI), underscores the necessity of ensemble learning, and elucidates the challenges inherent in evaluating these methodologies within the context of network intrusion detection tasks. 

\subsection{Types of Network Intrusions}

Various network intrusion types exist, categorized within the widely recognized MITRE ATT\&CK framework~\cite{strom2018mitre}. In our study, we address the primary network attacks outlined in this framework. Consequently, network traffic is broadly classified into the following categories: 

\textbf{Normal traffic:} This refers to regular network activity observed within the system.


\textbf{Malware / Malware Repository Information obtained regarding malicious software [MITRE ATT\&CK  ID: DS0004]:} This refers to analyzing malware for traits that might link it to specific creators, like the compiler used, debugging traces, code similarities, or group identifiers related to particular MaaS providers.  Finding overlaps in malware usage by different adversaries may suggest the malware was acquired rather than independently developed. In this context, overlapping features in malware used by various adversaries could indicate a shared quartermaster~\cite{Malware}.

\textbf{PortScan (PS) / Network Service Discovery [MITRE ATT\&CK  ID: T1046]:}  PortScan involves an intrusion where the attacker conducts reconnaissance on the victim's computer. Often utilized as an initial step in an attack, it aims to identify vulnerabilities and potential entry points. The method involves sending connection requests to various ports, without finalizing the connection. Responses received from these ports help map potential entry points for exploitation~\cite{lee2003detection}.

\textbf{Denial of Service (DoS) / Network Denial of Service [MITRE ATT\&CK ID: T1498]:} This type of attack aims to disrupt the target's network availability. A common example involves the attacker continuously sending connection requests to a server. However, upon receiving acknowledgment from the server, the attacker fails to respond, leaving the server's resources tied up and eventually leading to its unavailability. For comprehensive classifications of DoS attacks, readers are referred to~\cite{CICIDS}.

\textbf{Brute Force [MITRE ATT\&CK ID: T1110]:} This attack involves attempting all possible password combinations to gain unauthorized access to the victim's network. Attackers often leverage commonly used passwords in conjunction with this method. Success is more likely when users employ weak or easily guessable passwords~\cite{CICIDS}.

\textbf{Web Attack / Initial Access [MITRE ATT\&CK ID: TA0001, T1659, T1189]:} This category encompasses attacks conducted through web channels, exploiting vulnerabilities in web systems. For instance, attackers may exploit vulnerabilities in public-facing applications, leveraging software bugs, misconfigurations, or glitches to gain access to the application's underlying instance or container. Examples of such attacks include Drive-by Compromise~\cite{Web_attack_Ref2}. However, it is noteworthy that while web attacks such as SQL injection (SQLi) and Cross-Site Scripting (XSS) are common, they typically do not directly provide initial access to a remote server~\cite{strom2018mitre}.

\textbf{Infiltration / Initial Access [MITRE ATT\&CK ID: TA0001]:} This type of attack occurs when an unauthorized entity attempts to gain initial access to a system or application. It encompasses various techniques, including targeted spear phishing and exploiting vulnerabilities in public-facing web servers. The initial access gained through this attack can vary, ranging from simply changing a password to maintaining persistent access through legitimate accounts and external remote services.

\textbf{Botnet / Compromise Infrastructure [MITRE ATT\&CK ID: T1584.005, T1059, T1036, T1070]:} This type of attack involves the use of automated scripts executed remotely by attackers through hijacked devices. These scripts, known as bots, emulate human behavior and replicate it across multiple devices. The scripted nature of this technique enables scalability and easy deployment, making it an effective tool for targeting multiple attack points simultaneously. Consequently, botnets are a prevalent type of network attack.

\textbf{Probe Attack / Network Scanning or Surveillance [MITRE ATT\&CK ID: T1595]:} Probe attacks serve as the initial phase of a broader attack strategy. These attacks involve scanning a network to collect information or identify known vulnerabilities~\cite{chen2022intrusion}. Armed with a map detailing the available machines and services within a network, attackers can leverage this information to seek out potential exploits. It is important to note that while port scanning represents a type of probe attack, not all probe attacks involve port scans. Some may target specific vulnerabilities or utilize alternative methods, such as ping sweeps~\cite{gorodetski2002attacks} or DNS zone transfers~\cite{skwarek2019characterizing}.

\subsection{Intrusion Detection Systems}

The escalating complexity of cyber attacks poses a substantial risk to critical infrastructure across diverse industries~\cite{khan2021m2mon,hussain2021noncompliance}. As a result, IDS plays a pivotal role in defending computer network systems against malicious activities, whether perpetrated by internal users or external adversaries~\cite{mirzaei2021scrutinizer}. Conventional IDS architectures typically operate under the assumption that an intruder's actions will noticeably diverge from those of a legitimate user, thereby enabling the detection of many unauthorized activities~\cite{lukacs2015strongly}. With recent strides in artificial intelligence (AI) over the past decade, this architectural paradigm has facilitated the emergence of AI models capable of autonomously identifying network intrusions~\cite{kim2020ai}.

\subsection{Limitations of Base Learner Models}

While AI models have greatly automated intrusion detection, their inherent complexity presents constraints due to the intricate nature of their learning and decision-making mechanisms. This complexity poses challenges for a single model to fully grasp the subtleties of datasets, resulting in difficulties in learning specific subsets and achieving satisfactory metrics for certain outcomes. This challenge is widespread across various AI models, such as Decision Trees (DT), K-nearest neighbors (KNN), Support Vector Machines (SVM), Deep Neural Networks (DNN), among others. Despite their high predictive accuracy in Intrusion Detection Systems (IDS), there persists a gap in attaining better accuracy, precision, recall, and F1 scores, particularly in error or attack scenarios (including a high false positive rate for some AI models~\cite{AI_False_Rates} and a high false negative rate for others~\cite{mijalkovic2022reducing}). This issue is especially critical in safety-sensitive applications like network security through IDS. Consequently, there is a growing impetus to enhance performance and broaden the application of AI models in IDS. This has spurred the urgent need to employ diverse ensemble learning techniques to bolster IDS by leveraging the combination of different base learner models~\cite{al2021classification,aburomman2017survey,tama2021ensemble}. 

\subsection{Key Advantages of Ensemble Methods}


It is crucial to recognize that individual base learners possess distinct strengths and weaknesses. Depending on the specific application or task, one model may outperform others, adding complexity to the model selection process. Machine learning algorithms operate on diverse underlying principles. For instance, K-nearest neighbors (KNN), which clusters similar data around centroids, is sensitive to factors like the number of clusters ($K$), class outliers, and irrelevant features, besides being computationally demanding. Neural Networks (NN), on the other hand, typically require large datasets and substantial computational resources, while also being susceptible to variations in input data. Regression methods like Logistic Regression offer simplicity and interpretability but may struggle to capture intricate relationships, such as higher-order polynomials. Similarly, Decision Trees boast quick training times but can oversimplify problems, potentially leading to overfitting. Consequently, amalgamating these AI models through ensemble techniques can enhance their robustness, generalizability, and effectiveness in network intrusion detection tasks by leveraging their complementary strengths and mitigating their weaknesses.

Ensemble Learning is a dynamic field that delves into the concept of harnessing the strengths of diverse base learners to enhance predictive performance. Among the most renowned ensemble techniques are Bagging, Boosting, Blending, and Stacking. Bagging, short for Bootstrap Aggregating, involves creating multiple subsets of the dataset through bootstrapping, wherein data points are sampled with replacement, and training separate instances of a machine learning model on each subset. These models are trained independently. The primary objective of Bagging is to mitigate overfitting and enhance generalization by leveraging the diversity among the models. In contrast, Boosting operates by sequentially training multiple instances of the same base model, with each subsequent model aiming to correct the errors made by its predecessors. Boosting achieves this by assigning higher weights or emphasis to misclassified data points, effectively prioritizing instances that were previously difficult to classify correctly. By iteratively refining the model's performance, Boosting endeavors to improve predictive accuracy and reduce bias in the final ensemble. Meanwhile, the Stacking method adopts a distinct approach by training a diverse array of base learners and utilizing their predictions as features to train a meta-model. This meta-model learns to combine the predictions of the base learners, effectively capturing complex relationships between features and the target variable.

The ensemble methods such as Bagging, Boosting, and Stacking offer sophisticated strategies for improving predictive performance by leveraging the collective intelligence of diverse base learners. By combining the strengths of individual models and mitigating their weaknesses, ensemble techniques pave the way for more accurate and robust predictions across a wide range of machine-learning tasks.

In our study, we investigate various ensemble learning approaches within our framework, exclusively utilizing base models for network intrusion detection tasks. This comparative analysis is conducted across two distinct datasets, each possessing unique characteristics, in order to gain a comprehensive insight into our proposed framework.

%% file: Framework.tex

\section{Framework}\label{sec:framwork}

\begin{figure*}[t]
   \centering
\includegraphics[width= 0.7\linewidth]{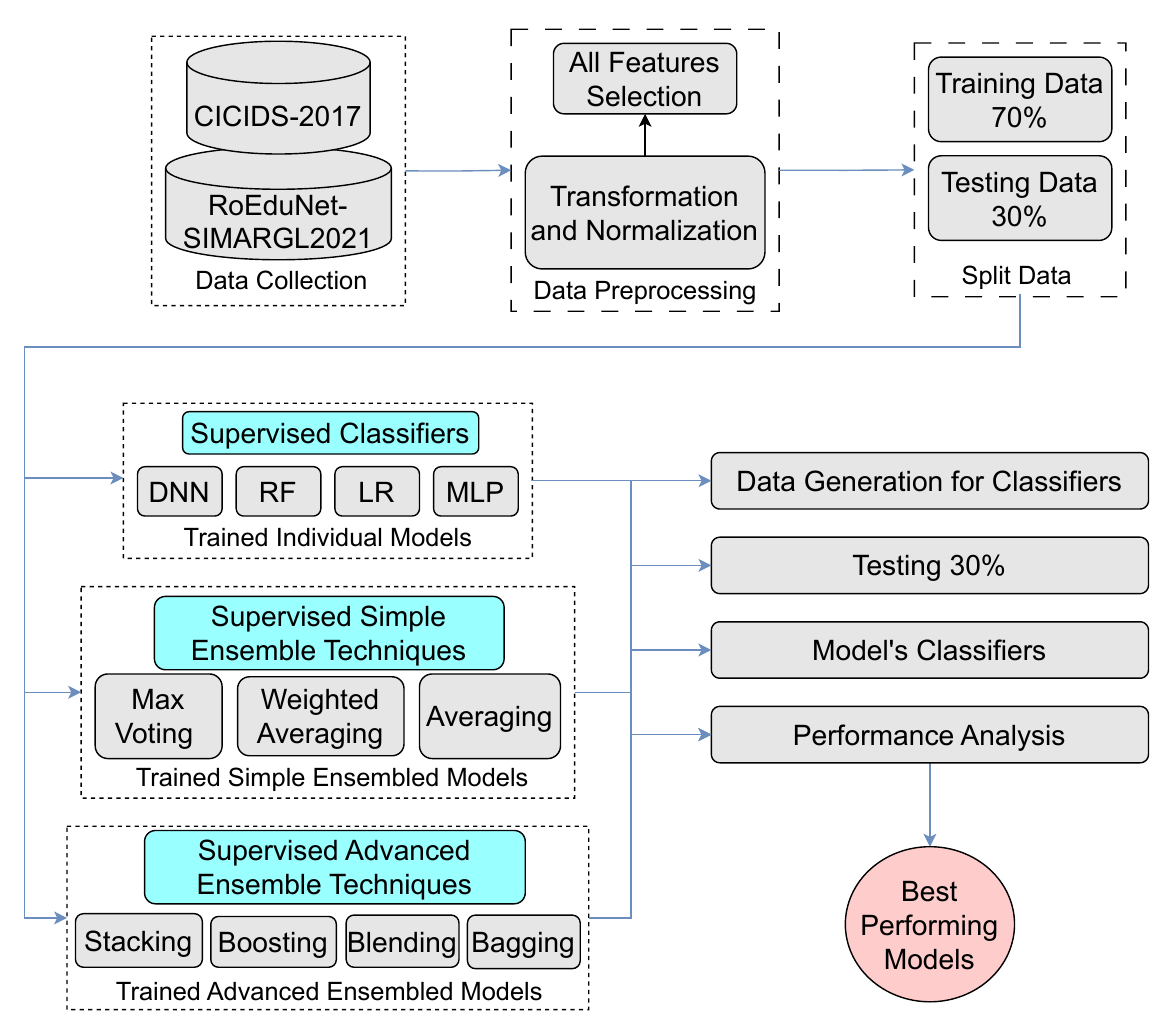}
   \caption{An overview of our ensemble learning framework for network IDS. It considers a diverse set of AI models and ensemble methods, along with network intrusion datasets.}
   \label{fig:systemframework}
\end{figure*}

The primary objective of this study is to develop an ensemble learning pipeline aimed at enhancing result metrics across diverse datasets. Our framework aims to assist security analysts in selecting effective methodologies for identifying intrusions and classifying attacks on network traffic, thereby bolstering intrusion prevention measures within their scope. To achieve this, we delineate a methodological framework comprising key stages for investigating the efficacy of various ensemble learning techniques tailored for intrusion detection systems (IDS), shown in Figure~\ref{fig:systemframework}.

\subsection{Data Preprocessing}
The CICIDS-2017 and RoEduNet-SIMARGL2021 datasets underwent thorough preprocessing for intrusion detection systems (IDS). For CICIDS-2017, duplicate records were removed, and missing values were imputed with the mean for the `Flow Bytes/s' column. Leading space characters in feature names were also removed, and label encoding was applied to categorical data in the `Label' column. 

Similarly, for RoEduNet-SIMARGL2021, duplicate records were removed, columns with singular unique values were dropped, and missing values were filled with the mean values for respective columns. Categorical features were encoded into numerical values using the Ordinal Encoder. These preprocessing steps aimed to enhance data quality and consistency for subsequent analyses.

\subsection{Model Selection and Used Techniques }
In this section, we will explain the model selection process including the selection of individual base learners and the use of ensemble techniques such as simple  and advanced ensemble techniques.

\subsubsection{The Individual Base Learner Models}  
We carefully selected a set of diverse and established base learners to leverage their complementary strengths. 
\begin{itemize}
    \item \textbf{Decision Trees:} Decision trees are well-known for their simplicity and easy-to-understand nature. They provide an intuitive representation of the decision-making processes in the data. 

    \item \textbf{Neural Networks:} We focused on the intricacy and non-linearity of neural networks, especially the Multi-layer perceptron Classifier.     
    \item \textbf{Logistic Regression:} It is a benchmark model that provides insights into linear relationships in the data.
\end{itemize}

\subsubsection{The Simple Ensemble Techniques}  
In conjunction with individual base learners, we employed various simple ensemble techniques to enhance predictive performance. These techniques included: 
\begin{itemize}
    \item \textbf{Averaging Predictions:} By averaging the predictions of multiple individual models, we aimed to reduce variance and improve overall prediction accuracy. 
    \item \textbf{Max Voting:} Employing a majority voting scheme, max voting aggregates predictions from multiple models and selects the most frequently occurring class label as the final prediction. 
    \item \textbf{Weighted Averaging:} Assigning weights to predictions from individual models based on their performance, weighted averaging allowed us to emphasize the contributions of more accurate models while mitigating the impact of less accurate ones. We explain how the weights are assigned in our experiments in the Evaluation Section. 

\end{itemize}

\subsubsection{The Advanced Ensemble Techniques}  
To further bolster our ensemble model's efficacy, we delved into advanced ensemble techniques, comprising:

\begin{itemize}
    \item \textbf{Bagging:} Through bootstrap aggregating, bagging generates diverse subsets of the training data and trains multiple base learners on each subset. By averaging their predictions, bagging reduces variance and enhances model robustness.  In this context, Random forests aggregate predictions from several decision trees to reduce overfitting and maintain robust predictive performance across different datasets. 
    
    \item \textbf{Blending:} Leveraging the outputs of multiple base learners as features, blending combines their predictions using a meta-learner to generate the final prediction. This technique harnesses the diversity of base learners to improve generalization.
    
    \item \textbf{Boosting:} Sequentially training base learners to correct the errors of preceding models, boosting emphasizes the misclassified instances, thereby iteratively refining the model's predictive performance.
    
    \item \textbf{Stacking:} Combining predictions from multiple base learners as features, stacking employs a meta-learner to learn the optimal combination of base learner predictions. This hierarchical ensemble technique leverages the diverse strengths of individual models to improve overall performance. 
\end{itemize}

\subsection{Model Implementation and Training} 

Following the meticulous selection of models, the implementation phase commenced, leveraging Python for the realization of our ensemble framework. Our implementation strategy began with the deployment of individual models, followed by the integration of simple ensemble techniques, culminating in the incorporation of diverse array of advanced ensemble techniques.


In order to make the best use of our high-performance system, we decided to use TensorFlow's distribution strategy, specifically tf.distribute.MirroredStrategy(). This strategy is designed for synchronous training across multiple GPUs within a single machine. It works by replicating the model's variables and computations across all available GPUs, which makes parallelism more efficient and speeds up the training process significantly. Each GPU independently computes gradients for a subset of the training data, and these gradients are aggregated across all GPUs to update the model's parameters. By synchronizing training across all GPUs, this approach maximizes GPU utilization, prevents inconsistencies, and ultimately accelerates the training process while improving overall efficiency. This strategy aligns perfectly with our goal of using our high-performance computer's computational resources to speed up model development and experimentation. 


\subsubsection{Individual Model Implementation and Training }
We implemented and trained each chosen base learner on its own, giving us the opportunity to explore its algorithms and performance characteristics in detail. With the help of Python’s powerful libraries (scikit-learn), TensorFlow, Keras, we were able to implement and train decision trees and random forests as well as neural networks (especially Multi-layer Perceptron Classifier), as well as logistic regression models with ease. We implemented and trained the decision trees model, we utilized the scikit-learn library. Decision Tree Classifier (DT) function was used to create a decision tree classifier object, and then we trained the classifier using the fit function with our training data. Similar approaches were followed for implementing and training other models, such as random forests, Multi-layer Perceptron Classifier, and logistic regression. Each model was instantiated using its respective class from scikit-learn, and then trained on training data using the ``fit'' function.

After training each model, we utilized the ``predict'' function to test the trained models using the test dataset that we prepared. This allowed us to evaluate the performance of each model on unseen data.

We further evaluated the models by computing their accuracy using the ``accuracy\_score'' function from scikit-learn. Additionally, we printed the classification report using the classification\_report function to obtain precision, recall, and F1-score for each class. We then visualized the performance of the models using confusion matrices generated by the confusion\_matrix function. These evaluations provided us with insights into the overall performance and effectiveness of each model for the task.

\subsubsection{Simple Ensemble Techniques Implementation and Training }
Following the individual model implementations, simple ensemble techniques were employed to combine predictions from multiple models, thereby enhancing predictive performance. Python's array manipulation capabilities and built-in functions facilitated the seamless integration of averaging, max voting, and weighted averaging techniques.

In the averaging technique implementation, three diverse classifiers (Decision Tree (DT), K-Nearest Neighbors (KNN), and Random Forest (RF)) are trained simultaneously within the distributed scope, enabling efficient GPU acceleration. Predictions are then made using each model, and a Soft Voting ensemble technique is applied to combine the predictions into a final output. This ensemble approach aims to enhance predictive performance by leveraging diverse models and distributed computing resources.

After training the ensemble model using averaging technique, we utilized the ``predict'' function to test the model using the test dataset. We further evaluated the model's performance by computing its accuracy using the ``accuracy\_score'' function and printing the classification report using the ``classification\_report'' function. Additionally, we generated a confusion matrix using the confusion\_matrix function to visualize the model's performance (as will be shown in next sections).

In the max voting technique implementation, a distributed training approach using TensorFlow's MirroredStrategy is employed to optimize the training process across multiple GPUs. Three distinct classifiers (K-Nearest Neighbors (KNN), Decision Tree (DT), and Random Forest (RF)) are initialized. These models are then integrated into a Voting Classifier using hard voting. The Voting Classifier aggregates the predictions from individual models and selects the class label with the majority vote as the final prediction. Subsequently, the Voting Classifier is trained on the training data, and its performance is evaluated using the test data. This ensemble approach enhances the model's predictive capability by leveraging  strengths of multiple classifiers.


In the weighted averaging technique implementation, a distributed training strategy using TensorFlow's MirroredStrategy is employed, facilitating parallel execution across multiple GPUs. Within this distributed scope, three distinct classifiers (Decision Tree (DT), K-Nearest Neighbors (KNN), and Random Forest (RF)) are instantiated. These classifiers are then integrated into a Voting Classifier, which aggregates their predictions using hard voting. Custom weights are assigned to each classifier to influence their contribution to the final prediction, with DT accounting for 40\%, KNN for 30\%, and RF for 30\%. Finally, the ensemble model is trained on the provided training data. This ensemble-based approach aims to enhance predictive accuracy by leveraging the diverse capabilities of individual classifiers while considering their respective contributions to the final prediction.

Following the training of the ensemble model using weighted averaging technique, we conducted prediction and evaluation by computing accuracy, printing the classification report, and generating the confusion matrix.

\subsubsection{Advanced Ensemble Techniques Implementation}
Finally, advanced ensemble techniques were implemented to further enhance the predictive capabilities of the model. Python's machine learning libraries, including scikit-learn, provide seamless integration of various advanced ensemble techniques such as bagging, blending, boosting (including Adaptive Boosting, Cat Boosting, Gradient Boosting, and XGBoost Extreme Gradient Boosting), and stacking.

For the bagging technique implementation, a distributed training strategy using TensorFlow's MirroredStrategy is initiated to enable parallel execution across multiple GPUs. Within this distributed context, a list of diverse base models is instantiated, including RF, MLP, LR, and DT classifiers. These base models serve as the foundational components for the ensemble approach. Subsequently, a Bagging Classifier is constructed, utilizing RF as the base model. Bagging is a robust ensemble technique effective in reducing overfitting by aggregating predictions from multiple models. In this implementation, the Bagging Classifier is configured with the same number of estimators as the number of base models to ensure diversity and effectiveness in prediction. The Bagging Classifier is then trained on the provided training data.


Similarly, for the blending technique implementation, TensorFlow's MirroredStrategy is employed to facilitate parallel execution across multiple GPUs. Under this strategy's scope, several base models including RF, MLP, LR, and DT are initialized and trained. Each of these models generates predictions for the test data, which are then combined using the blending technique to create a new dataset. This dataset serves as the input for a meta-model, another DT Classifier. The meta-model is trained on the blended predictions to learn how to best combine the outputs of the base models. Additionally, predictions are made on the test set using the trained base models, and a new dataset (blend\_X\_test) is created with these predictions to be used as input for the meta-model. Finally, the meta-model predicts the final output based on the blended predictions from the base models.


Along the same lines, for the boosting technique implementations including Adaptive Boosting, Cat Boosting, Gradient Boosting, and XGBoost Extreme Gradient Boosting, we followed the same aforementioned process.

For the stacking technique implementation, TensorFlow's MirroredStrategy is employed to facilitate parallel execution across multiple GPUs. Within this distributed context, four base models (RF, MLP, LR, and DT) are instantiated and integrated into pipelines incorporating Principal Component Analysis (PCA) for dimensionality reduction. These pipelines preprocess the data, enhancing model performance and reducing computational complexity. A meta-model, represented by another DT Classifier, is instantiated to learn from the outputs of the base models. The Stacking Classifier from scikit-learn is utilized to stack the base models, combining their predictions as features for the meta-model. Finally, the stacked model is trained on the provided training data, enabling it to learn the optimal combination of predictions from the base models.

\subsection{Evaluation Metrics and Model Selection Rationale}

\textbf{Results' Metrics:} To evaluate the performance of the selected models and techniques comprehensively, we employed four primary performance indicators: Accuracy, Precision, Recall, and F1 score. Additionally, runtime was considered as a metric to assess the computational efficiency of the models. These metrics collectively provide insights into the effectiveness and efficiency of the models in detecting intrusions.

We organized the results systematically to facilitate analysis and comparison across different models and techniques. This structured approach enables us to draw meaningful conclusions regarding the suitability and efficacy of the models for IDS applications.

\textbf{Model Selection Criteria:} The models chosen for this study were selected based on several key factors. Primarily, their prevalence in prior research pertaining to Intrusion Detection Systems (IDS) ensured alignment with established literature, enabling effective comparison with seminal studies such as~\cite{CICIDS, dhanabal2015study}. Furthermore, these diverse ensemble learning methods had success in different applications. By adopting widely-used models, our research maintains consistency with existing methodologies, facilitating a robust evaluation of various models and ensemble learning techniques utilized in our investigation. In this context, we emphasize that we used different AI models with different working principles for our ensemble learning (i.e., the KNN uses a different reasoning from MLP that also uses a different reasoning than DT).

\subsection{Comprehensive Overview of Top Network Intrusion Features and Their Role in the Learning}

In this subsection, we present a detailed list of the top network intrusion features along with their explanations for the two datasets under study, as they play a crucial role throughout the entirety of our paper. Tables~\ref{tbl:feature_list_sensor} and~\ref{tbl:feature_list_CICIDs} provide descriptions for each feature in the RoEduNet-SIMARGL2021 and CICIDS-2017 network intrusion datasets, respectively. 

Tables~{\ref{tbl:feature_list_sensor} and \ref{tbl:feature_list_CICIDs}} elucidate key features specific to the RoEduNet-SIMARGL2021 and CICIDS-2017 datasets. These tables serve to highlight significant features from each dataset, offering clarity and contextual understanding. However, it is essential to clarify that all features listed in Table~{\ref{tbl:samples_distributions_datasets}} were utilized in our preliminary experiments. This inclusive approach enabled us to fully exploit the datasets for our analysis of network intrusion detection. Notably, Table~{\ref{tbl:samples_distributions_datasets}} summarizes the overall composition of each dataset, encompassing the number of features.

\begin{table}[ht]
\centering
\caption{Description of  main features for RoEduNet-SIMARGL2021 dataset \cite{flow1234}.} 
\resizebox{\columnwidth}{!}{
\begin{tabular}{l|l}
\hline
\textbf{RoEduNet-SIMARGL2021 Features}               & \textbf{Explanation}      \\                               \hline                                     FLOW\_DURATION\_MILLISECONDS & Flow duration in milliseconds                         \\
PROTOCOL\_MAP                                                           & IP protocol name (tcp, ipv6, udp, icmp)                                                                   \\
TCP\_FLAGS                                                              & Cumulation of all flow TCP flags \\

TCP\_WIN\_MAX\_IN            & Max TCP Window (src-\textgreater{}dst)       \\

TCP\_WIN\_MAX\_OUT           & Max TCP Window (dst-\textgreater{}src) \\

TCP\_WIN\_MIN\_IN            & Min TCP Window (src-\textgreater{}dst) \\

TCP\_WIN\_MIN\_OUT   & Min TCP Window (dst-\textgreater{}src)       \\

TCP\_WIN\_SCALE\_IN  & TCP Window Scale (src-\textgreater{}dst)   \\

TCP\_WIN\_MSS\_IN  & TCP Max Segment Size (src-\textgreater{}dst) \\

TCP\_WIN\_SCALE\_OUT & TCP Window Scale (dst-\textgreater{}src)   \\
SRC\_TOS                                                                & TOS/DSCP (src-\textgreater{}dst)             \\
DST\_TOS                                                                & TOS/DSCP (dst-\textgreater{}src)  \\

FIRST\_SWITCHED & SysUptime of First Flow Packet\\

LAST\_SWITCHED & SysUptime of Last Flow Packet\\
TOTAL\_FLOWS\_EXP & Total number of exported flows
\end{tabular}
}
\label{tbl:feature_list_sensor}
\end{table}

\begin{table}[h]
\centering
\vspace{1mm}
\caption{Description of the main features for the CICIDS-2017 dataset~\cite{ahlashkari_2021}.}
\resizebox{\columnwidth}{!}{
\begin{tabular}{l|l}
\hline
\begin{tabular}[c]{@{}c@{}}\textbf{CICIDS-2017  Features}\end{tabular} & \textbf{Explanation}                                                                                                   \\ \hline
Packet Length Std                                         & Standard deviation  length of a packet                                      \\
Total Length of Bwd Packets                               & Total size of packet in backward direction                          \\
Subflow Bwd Bytes                                         & Average number of bytes in backward sub-flow        \\
Destination Port                                          & Destination Port Address                                  \\
Packet Length Variance                                    & Variance length of a packet                               \\
Bwd Packet Length Mean                                    & Mean size of packet in backward direction                           \\
Avg Bwd Segment Size                                      & Average size observed in the backward direction                     \\
Bwd Packet Length Max                                     & Maximum size of packet in backward direction                     \\
Init\_Win\_Bytes\_Backward                                & Total number of bytes in initial backward window \\
Total Length of Fwd Packets                               & Total packets in the forward direction                                                                                \\
Subflow Fwd Bytes                                         & Average number of bytes in a forward sub-flow  \\
Init\_Win\_Bytes\_Forward                                 & Total number of bytes in initial forward window \\
Average Packet Size                                       & Average size of packet                                                                                                \\
Packet Length Mean                                        & Mean length of a packet                                                                                               \\
Max Packet Length                                         & Maximum length of a packet                                    
\end{tabular}
}
\label{tbl:feature_list_CICIDs}
\end{table}

\begin{table}[h]
\caption{Summary and statistics of the three network intrusion datasets used in this work, including the size of the dataset, number of attack types (labels), number of intrusion features, and distribution of samples among attack types.}
\label{tbl:combined_dataset_statistics}

\begin{subtable}{0.48\textwidth}
\label{tbl:basic_dataset_statistics}
\caption{Basic statistics of datasets}

\resizebox{\textwidth}{!}{

\begin{tabular}{|l|c|c|c|}
\hline
\textbf{Dataset} & \textbf{No. of Labels} & \textbf{No. of Features} & \textbf{No. of Samples} \\ \hline
\textbf{CICIDS-2017} & 7 & 78 & 2,775,364 \\ \hline
\textbf{RoEduNet-SIMARGL2021} & 3 & 29 & 31,433,875 \\ \hline
\end{tabular}
}
\end{subtable}

\vspace{0.6em}

\begin{subtable}{0.48\textwidth}
\caption{Distribution of samples among different attack types}
\label{tbl:samples_distributions_datasets}
\resizebox{\textwidth}{!}{
\begin{tabular}{|l|c|c|c|c|c|c|c|}
\hline
\textbf{Dataset} & \textbf{Normal} & \textbf{DoS} & \textbf{PortScan} & \textbf{Brute Force} & \textbf{Web Attack} & \textbf{Bot} & \textbf{Infiltration} \\ \hline
\textbf{CICIDS-2017} & 84.442\% & 9.104\%  & 5.726\%  & 0.498\% & 0.157\% & 0.071\% & 0.001\% \\ \hline
\textbf{RoEduNet2021} & 62.20\% & 24.53\% & 13.27\% & - & - & - & -  \\ \hline

\end{tabular}
}
\end{subtable}

\end{table}

%% file: Evaluation.tex

\section{Foundations of Evaluation}\label{sec:evaluation}

In this section, we present a comprehensive evaluation aimed at addressing key research questions that underpin our study:

\begin{enumerate}
\item What are the optimal individual ML models suited for a given network intrusion detection dataset?
\item Which ensemble method exhibits superior performance on a given dataset?
\item How do the evaluated methods within our framework perform across key metrics such as Accuracy, Precision, Recall, F1 Score, and runtime?

\item What are the inherent limitations and strengths associated with the application of ensemble learning methods in the context of network intrusion detection?
\end{enumerate}

Before delving into the detailed evaluation results, we provide a comprehensive overview of the experimental setup.

\subsection{DataSet Description}

\textbf{RoEduNet-SIMARGL2021 Dataset~\cite{mihailescu2021proposition}:} This dataset stems from the SIMARGL project, a collaborative initiative supported by the European Union under the Horizon program, in conjunction with the Romanian Education Network (RoEduNet). It comprises authentic network traffic data, incorporating features derived from real-time traffic analysis. The dataset adheres to a structured data schema reminiscent of Netflow~\cite{claise2004cisco}, a network protocol developed by CISCO for the purpose of capturing and monitoring network flows.

\textbf{CICIDS-2017 Dataset~\cite{panigrahi2018detailed}:} Serving as a benchmark for intrusion detection, this dataset was curated by the Canadian Institute for Cybersecurity at the University of Brunswick in 2017. It encompasses six distinct attack profiles, including activities such as brute force, heartbleed, botnet, Denial of Service (DoS), portscan, web attack, and infiltration attack. To establish a realistic context, the dataset incorporates background traffic generated through a B-Profile system~\cite{sharafaldin2018towards}, which captures various user behaviors based on popular network protocols.

\textbf{Summary and Statistics of the Datasets:} Each dataset is characterized by its size, the number of attack types (labels), and the quantity of intrusion features. Detailed statistics regarding these attributes are presented in Table~\ref{tbl:combined_dataset_statistics}.

\vspace{-2mm}
\subsection{Experimental Setup}

\textbf{Computing Resources:} Our experiments were conducted on a high-performance computing (HPC) system equipped with robust hardware capabilities. The HPC configuration includes two NVIDIA A100 GPUs, 64 GPU-accelerated nodes, each boasting 256 GB of memory, and a single 64-core AMD EPYC 7713 processor running at 2.0 GHz with a power consumption of 225 watts. This setup enables a peak performance of approximately 7 petaFLOPs, making it exceptionally well-suited for intensive AI and machine learning tasks~\cite{BIGRED200}.


\textbf{Coding Tools:} To ensure versatility and openness in our implementation, we utilized the Python programming language alongside various AI toolboxes such as Keras and ScikitLearn. Additionally, we leveraged essential libraries including Pandas and Matplotlib. By adopting these open-source tools, we aimed to facilitate transparency and reproducibility in our research endeavors.

\subsection{Evaluation Metrics}

In this study, the utilization of well-established evaluation metrics is crucial to ascertain the most effective model for integration within an Intrusion Detection System (IDS). Accuracy, precision, recall, and F1-score stand as quintessential performance evaluation metrics. These metrics are derived from four fundamental measures: true positive (TP), false positive (FP), true negative (TN), and false negative (FN) rates. The evaluation metrics are delineated as follows:

\begin{itemize}
\item \textbf{Accuracy} $[(TP + TN)/Total]$: Signifies the proportion of accurately identified network traffic instances over the total data instances.

\item \textbf{Precision} $[TP/(FP + TP)]$: Measures the frequency with which the model accurately discerns an  attack.


\item \textbf{Recall} $[TP/(FN + TP)]$: Measures the model's ability to correctly identify  attacks (or intrusions). Recall is also referred to as the true-positive rate, sensitivity, or detection rate.

\item \textbf{F1-Score} $[2TP/(2TN + FP + FN)]$: Represents the harmonic mean of precision and recall.
\end{itemize}


\subsection{AI Models}

In this section, we outline the  main AI models employed in our study.

\textbf{(i) Base Learners:} We utilized four widely-used AI classification algorithms as base learners, namely: Multi-Layer Perceptron (MLP)~\cite{MEBAWONDU2020e00497}
Decision Tree (DT)~\cite{song2015decision}, Logistic Regression (LR)~\cite{dreiseitl2002logistic}, and
k-Nearest Neighbor (KNN)~\cite{li2014new}.  These AI methods form the foundation of our evaluation, allowing us to assess both their individual performances and their contributions to our network intrusion detection framework.

\textbf{(ii) Ensemble Methods:} In addition to the base learners, our framework incorporates advanced ensemble techniques such as stacking, blending, boosting (including Cat Boosting (CAT)\cite{dorogush2018catboost}, Light Gradient-Boosting Machine (LGBM)\cite{jin2020swiftids}, AdaBoost (ADA)\cite{yulianto2019improving}, Gradient Boosting (GB)\cite{natekin2013gradient}, Extreme Gradient-Boosting (XGBoost)~\cite{dhaliwal2018effective}), Random Forest, and bagging techniques. Furthermore, we employ simpler ensemble methods like Voting~\cite{dietterich2000ensemble}, Averaging~\cite{zounemat2021ensemble}, and Weighted Averaging, alongside the aforementioned models. These ensemble methods enhance the robustness and accuracy of our intrusion detection system.

\textbf{Hyperparameters :} We provide our main hyperparameter choices for each AI model and each ensemble method used in our work in Appendix~\ref{app:ai_models_hyperparams}.

Having provided the main experimental setup, we next detail our evaluation results and findings.


\begin{table}[t]
\centering
\vspace{1mm}
\caption{Performance of different models (both base learners and ensemble methods) on RoEduNet-SIMARGL2021 datasets. The results are organized by F1 score (highest to lowest).}
\vspace{-1mm}
 \resizebox{1\linewidth}{!}{
\begin{tabular}{|l|c|c|c|c|}
\hline
\textbf{Models} & \textbf{Accuracy (ACC)} & \textbf{Precision (PRE)} & \textbf{Recall (REC)} & \textbf{F1 Score} \\
\hline
Random Forest (RF)      & 1.00 & 1.00 & 1.00 & 1.00 \\
Decision Tree (DT)      & 1.00 & 1.00 & 1.00 & 1.00 \\
Average (avg)           & 1.00 & 1.00 & 1.00 & 1.00 \\
Max Voting (Max\_Vot) & 0.999 & 1.00 & 1.00 & 1.00  \\
Stacking                & 0.99998 & 1.00 & 1.00 & 1.00 \\
Weighted Average (weighed\_avg) & 0.998 & 1.00 & 1.00 & 1.00 \\
Bagging (Bag)           & 0.998 & 1.00 & 1.00 & 1.00 \\
Blending (Bled) & 0.998 & 1.00 & 1.00 & 1.00 \\
AdaBoost (ADA) & 0.99981 & 1.00 & 1.00 & 1.00 \\
Cat Boost (CAT) & 0.998 & 0.998 & 0.998 & 0.998 \\
Gradient Boosting (GB) & 0.988 & 0.99 & 0.99 & 0.998 \\
XGBoost (XGB) & 0.996 & 0.996 & 0.996 & 0.996 \\
Logistic Regression (LR) & 0.6781 & 0.56 & 0.68 & 0.58 \\
Multi-Layer Perceptron (MLP) & 0.6178 & 0.38 & 0.62 & 0.47 \\  
\hline
\end{tabular}
}
\label{table:SIMARGL2021_all_features}
\end{table}

\begin{table}[h]
\centering
\vspace{1mm}
\caption{Model Training and Testing Timetable (Seconds) for RoEduNet-SIMARGL2021 Dataset. The runtimes are organized (from shortest to lengthiest). Logistic Regression is the most efficient individual model while bagging is the most time-efficient ensemble method for this dataset.}
\vspace{-1mm}
\resizebox{0.9\linewidth}{!}{
\begin{tabular}{|l|c|}
\hline
\textbf{Models} & \textbf{Time (Seconds)} \\
\hline
Logistic Regression (LR) & 122.72 \\
Decision Tree (DT) & 900.74 \\
Multi-Layer Perceptron (MLP) & 6200.45 \\
Bagging (Bag) & 9834.9 \\
Random Forest (RF) & 10800.85 \\
XGBoost (XGB) & 12520.23 \\
AdaBoost (ADA) & 12610.54 \\
Cat Boost (CAT) & 18050.12 \\
Gradient Boosting (GB) & 18250.02 \\
Blending (Bled) & 19600.43 \\
Average (avg) & 27216.16 \\
Stacking & 28800.32 \\
Weighted Average (weighed\_avg) & 30816.53 \\
Max Voting (Max\_Vot) & 32400.452 \\
\hline
\end{tabular}
}
\label{table:SML_time_table}
\end{table}

\subsection{RoEduNet-SIMARGL2021 Analysis}

\textbf{Main Results :} Table~\ref{table:SIMARGL2021_all_features} presents the performance metrics of various models on the RoEduNet-SIMARGL2021 dataset when utilizing all available features. Notably, the majority of models demonstrate remarkable performance across all metrics, with particularly high values for precision, recall, and F1 scores. The top-performing techniques and models, including Random Forest (RF), Decision Tree (DT), Average (Avg), Max Voting (Max\_Vot), Stacking, and Bagging (Bag), consistently achieve perfect scores (1.00) across all metrics. This convergence suggests robust model performance across different ensemble techniques and underscores the efficacy of utilizing the complete feature set for classification.

Conversely, Logistic Regression (LR) and Multi-Layer Perceptron (MLP) exhibit comparatively lower performance metrics, indicating potential limitations in capturing the underlying patterns within the dataset.

Given the already optimal performance attained by several models, further optimization may yield marginal improvements at best. However, exploring alternative feature engineering strategies or investigating potential data augmentation techniques could offer avenues for enhancing model generalization and resilience to unseen data.

\textbf{Runtime Performance:} Table~\ref{table:SML_time_table} offers insights into the runtime performance of RoEduNet-SIMARGL2021 models, measured in seconds. Models such as Logistic Regression (LR) and Decision Tree (DT) demonstrate shorter runtimes, reflecting their computational efficiency. In contrast, ensemble methods and gradient boosting algorithms like Stacking, Average, and Gradient Boosting exhibit longer runtimes due to their inherent complexity and resource-intensive nature. However, their high performance across evaluation metrics justifies the computational investment, particularly in accuracy-sensitive applications.

Considering the extensive RoEduNet-SIMARGL2021 dataset comprising approximately 30 million samples, prioritizing models with optimal performance across metrics while balancing computational efficiency is crucial. Models like LR and DT emerge as promising candidates due to their shorter runtimes, making them attractive for scenarios with computational constraints. Conversely, models with longer runtimes, such as Blending (Bled) and Stacking, may require substantial resources. Nevertheless, their superior performance justifies resource allocation, especially in applications prioritizing precision and recall metrics.

To manage computational complexity for blending and stacking, experiments were conducted on a subset of the dataset, limiting the sample size to 20\% through random sampling, ensuring manageable overhead while retaining analytical integrity.

\subsubsection{The models and ensemble Confusion Matrices for RoEduNet-SIMARGL2021}

\begin{figure}
    \centering  \includegraphics[width= \linewidth]{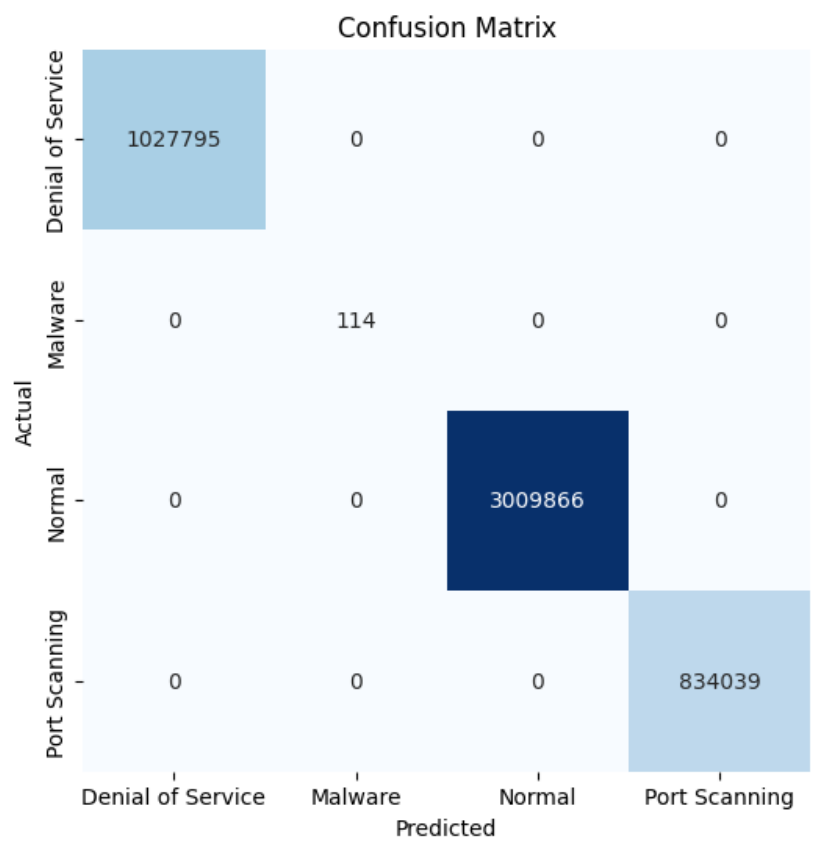}
\caption{Confusion Matrix of different models with perfect performance on RoEduNet-SIMARGL2021 Dataset. These models are: Random Forest (RF), Decision Trees (DT), averaging ensemble
technique (avg), Weighted Averaging (weighed\_avg), Bagging ensemble
technique (Bag), and Max Voting ensemble technique (Max\_Vot).}
    \label{fig:confusion-matrix-RF}
\end{figure}

The following results shown in this section provides the classification accuracy of the methods via confusion matrices that are displayed using a heat map. The 14 different learners are tested to classify four various attacks in the RoEduNet-SIMARGL2021 Dataset.  We group those as follows:

\textbf{(i) Confusion Matrices of Models with Perfect Performance:}  Figure~\ref{fig:confusion-matrix-RF} shows the confusion Matrix of different models with
perfect performance on RoEduNet-SIMARGL2021 Dataset. These models are: Random Forest (RF), Decision Trees (DT),
averaging ensemble technique (avg), Weighted Averaging
(weighed\_avg), Bagging ensemble technique (Bag), and Max
Voting ensemble technique (Max\_Vot). The figure shows that all normal samples along with the  three intrusion classes (Denial of Service, Malware, and Port Scanning) have been predicted perfectly for these models.


\begin{figure}
    \centering   \includegraphics[width=\linewidth]{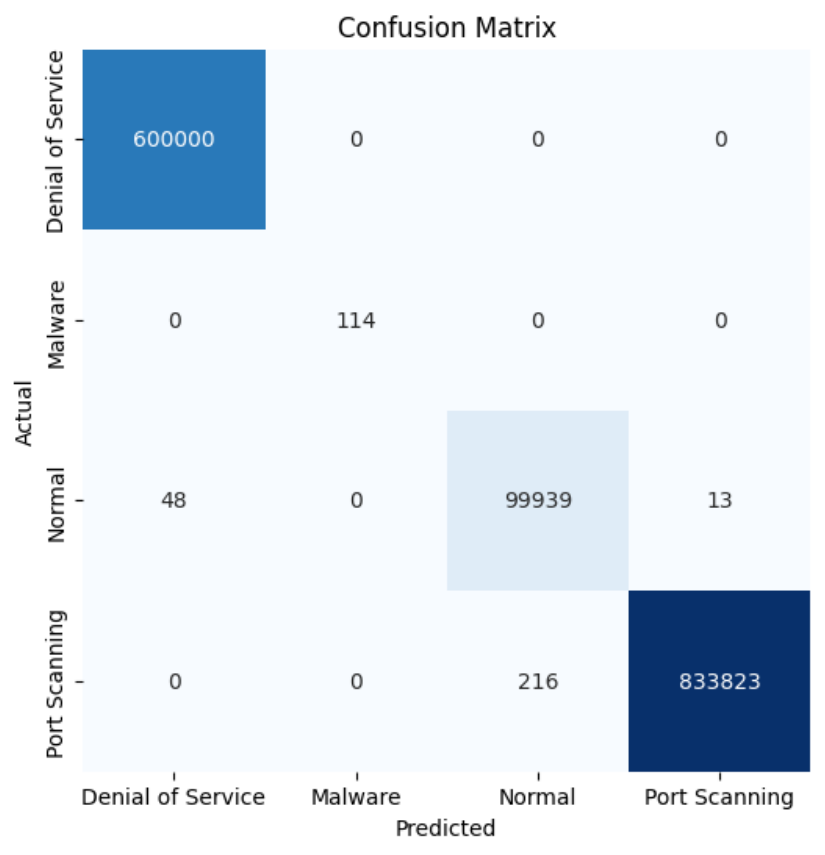}
    \caption{Confusion Matrix of Adaptive Boosting (ADA) ensemble technique on RoEduNet-SIMARGL2021 Dataset. It has near-perfect performance.}
    \label{fig:confusion-matrix-ADA}
\end{figure}

\begin{figure}
    \centering
    \includegraphics[width=\linewidth]{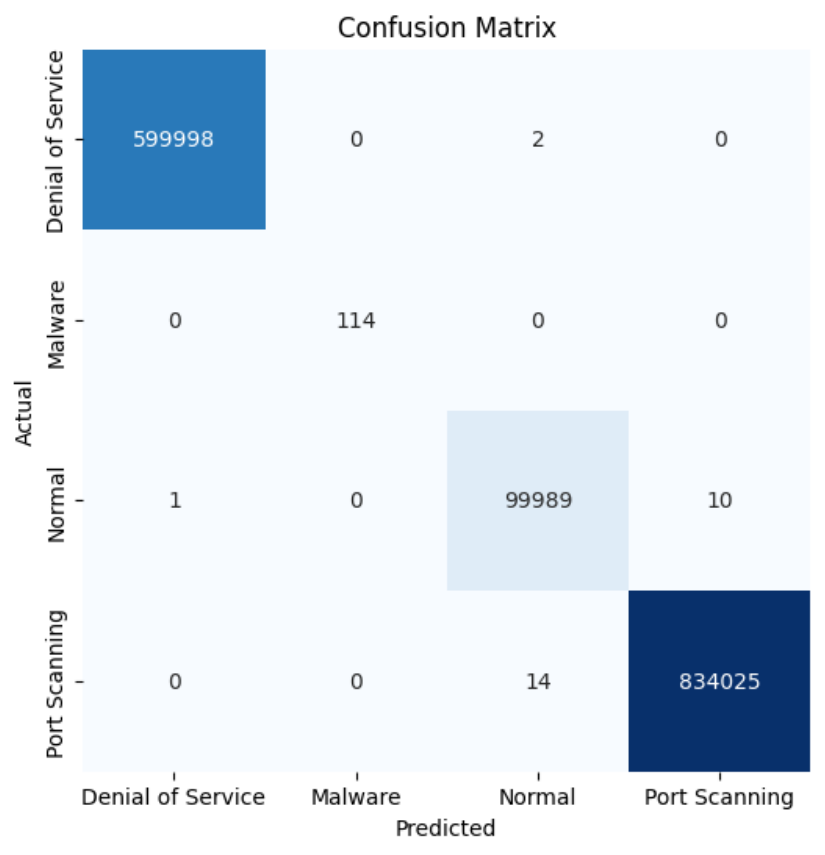}
    \caption{Confusion Matrix of Stacking ensemble technique on RoEduNet-SIMARGL2021 Dataset. It has near-perfect performance on the samples.}
    \label{fig:confusion-matrix-Stack}
\end{figure}

\begin{figure}
    \centering
    \includegraphics[width= \linewidth]{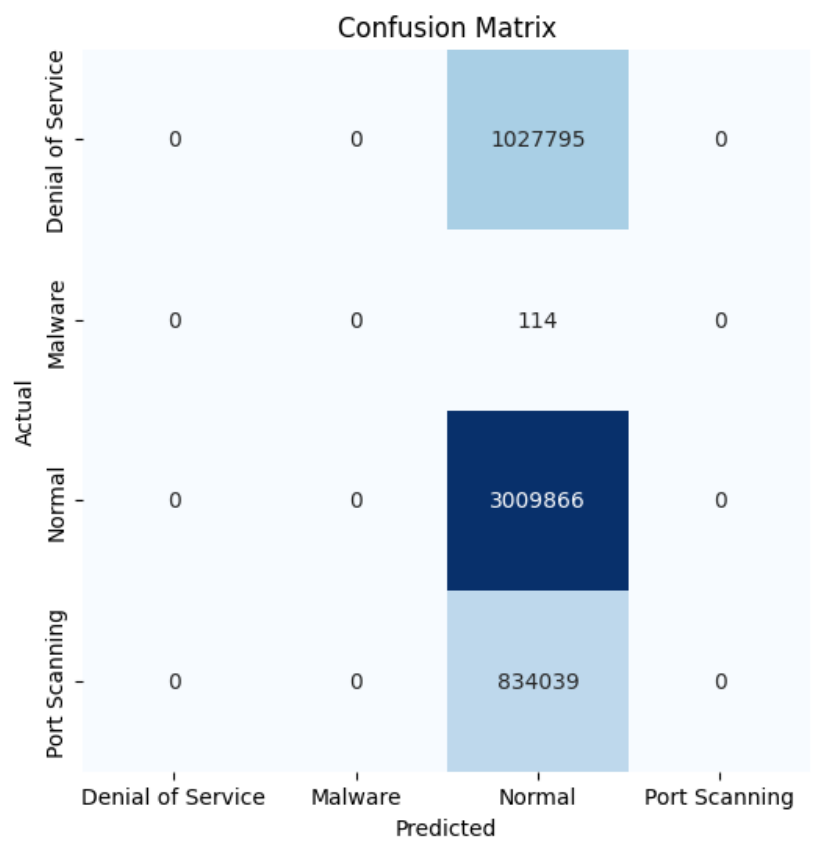}
    \caption{Confusion Matrix of MLP Classifier (MLP) on RoEduNet-SIMARGL2021 Dataset. It has low prediction performance, particularly for Denial of Service and Port Scanning attacks. On the other hand, it has perfect performance in identifying normal samples.}
\label{fig:confusion-matrix-MLPC}
\end{figure}

\begin{figure}
    \centering
    \includegraphics[width=\linewidth]{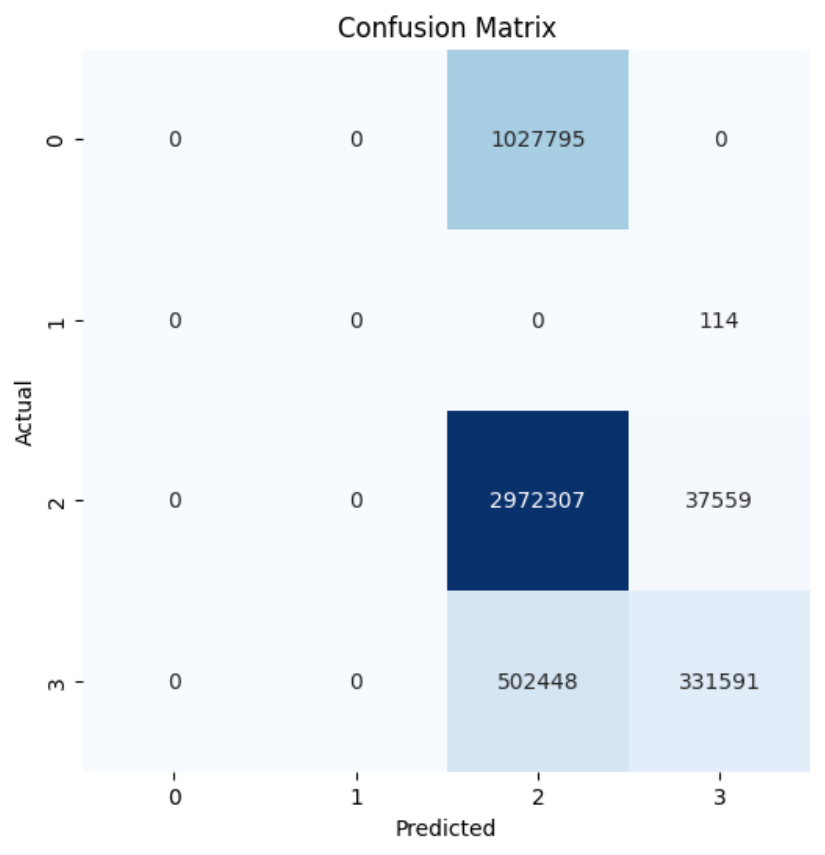}
    \caption{Confusion Matrix of Logistic Regression (LR) on RoEduNet-SIMARGL2021 Dataset. Here, label 0 represents Denial of Service (DOS), 1 represents Malware, 2 represents Normal, and 3 represents Port Scanning. LR has low performance on DoS and Port Scanning intrusions while having near-perfect performance on normal (label 2) samples.
 }
    \label{fig:confusion-matrix-lr}
\end{figure}

\textbf{(ii) Confusion Matrices of  Models with Near-Perfect Performance:} The second category of models include those that have near-perfect performances but have few errors in prediction. This category includes Adaptive Boosting (ADA) (Figure~\ref{fig:confusion-matrix-ADA}) and Stacking ensemble (Figure~\ref{fig:confusion-matrix-Stack}). 

\textbf{(iii) Confusion Matrices of Methods with Low Performance:} We finally show the confusion matrices for the models with the lowest performaces on our first dataset. Figure~\ref{fig:confusion-matrix-MLPC} shows the  confusion Matrix of MLP Classifier on RoEduNet-SIMARGL2021 Dataset. It has low prediction performance, particularly for Denial of Service and Port Scanning attacks. On the other hand, it has perfect performance in identifying normal samples. Moreover, Figure~\ref{fig:confusion-matrix-lr} shows the confusion Matrix of Logistic Regression (LR) on RoEduNet-SIMARGL2021 Dataset. LR has low performance on Denial of Service and Port Scanning intrusions while having near-perfect performance on normal traffic instances (or samples).

\begin{figure}
    \centering
    \includegraphics[width=\linewidth]{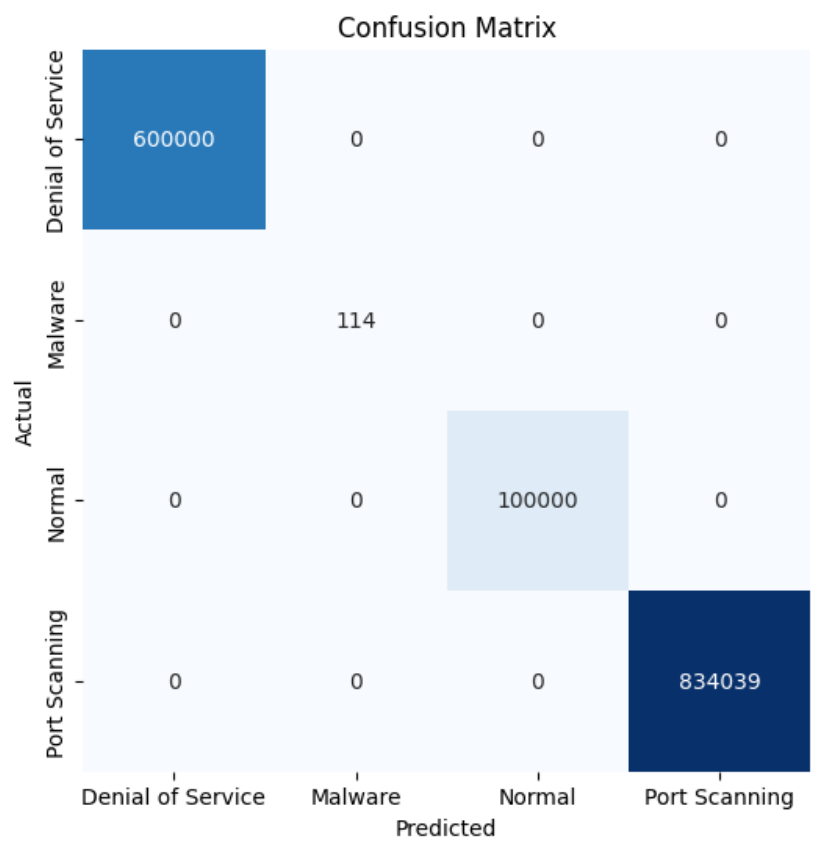}
    \caption{Confusion Matrix of Bagging (Bag), Blending (Bled), and Cat Boosting (CAT) ensemble techniques on RoEduNet-SIMARGL2021 Dataset.}
    \label{fig:confusion-matrix-Bag}
\end{figure}

\begin{figure}
    \centering
    \includegraphics[width=0.9\linewidth]{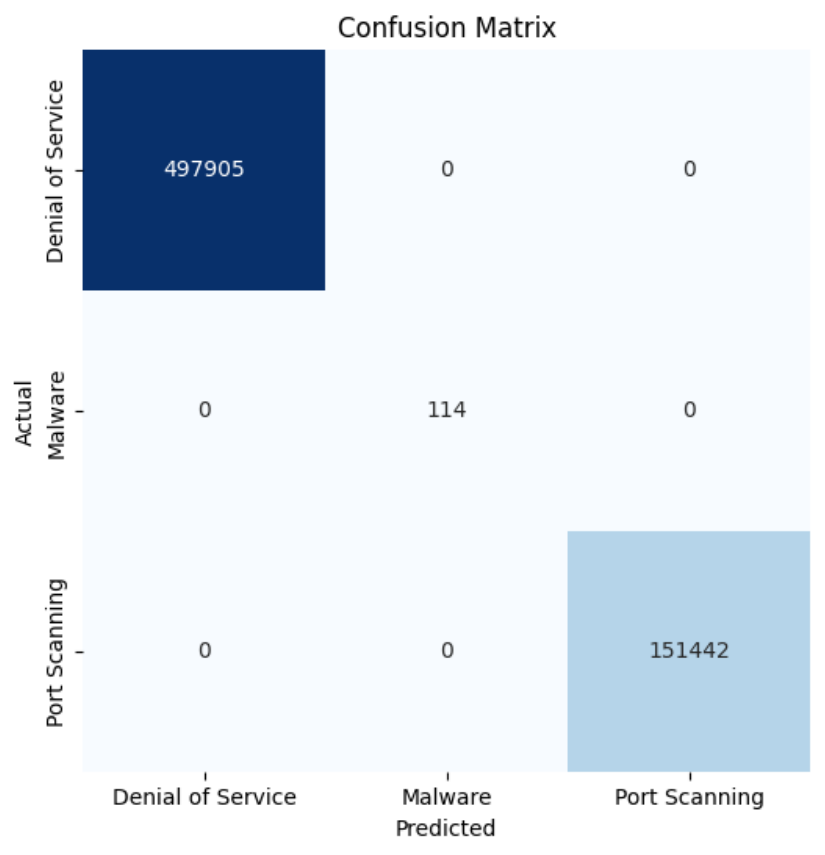}
    \caption{Confusion Matrix of Extreme Gradient Boosting (XGB) and Gradient Boosting (GB) ensemble techniques  on RoEduNet-SIMARGL2021 Dataset. Note that these two methods was tested on reduced sample size.}
    \label{fig:confusion-matrix-XGB}
    \vspace{-3mm}
\end{figure}

\textbf{(iv) Confusion Matrices for Boosting Models with Reduced Sample Size:} Finally, we show confusion matrices for the models that have computational issues on RoEduNet-SIMARGL2021 Dataset. Figure~\ref{fig:confusion-matrix-XGB} shows the confusion Matrices of these models, which are Extreme Gradient Boosting (XGB), and Gradient Boosting
(GB) ensemble techniques. Note that these two methods was tested on reduced sample size, however having almost perfect performances on all intrusion classes. Similarly, Figure~\ref{fig:confusion-matrix-Bag} shows the confusion Matrices for Bagging, Blending, and CAT boosting techniques. 

Having finished our detailed evaluation analysis on RoEduNet-SIMARGL2021 dataset, we next show the detailed evaluation analysis for CICIDS-2017 dataset.
    

\subsection{CICIDS-2017 Analysis}

\textbf{Main Results:} Table~\ref{table:cic2017_all_features} presents the performance metrics of various models on the CICIDS-2017 dataset when utilizing all available features. The table provides insights into the effectiveness of different machine learning algorithms in classifying network traffic. Notably, several models demonstrate high performance across all metrics, with Random Forest (RF), Bagging (Bag), Blending (Bled), Weighted Average (weighed\_avg), Stacking, Gradient Boosting (GB), Decision Tree (DT), and Cat Boost (CAT) consistently achieving near-perfect scores (F1 score of 0.998 or higher). This robust performance underscores the suitability of ensemble methods and tree-based algorithms for the classification task on this dataset.

Conversely, Logistic Regression (LR) and AdaBoost (ADA) exhibit comparatively lower performance metrics, suggesting potential limitations in capturing the complex patterns present in the dataset. The overlapping top-performing models further emphasize their stability and reliability across different feature sets.

\textbf{Runtime Performance:} Table~\ref{table:cic2017_time_table} presents the runtime performance of various machine learning models on the CICIDS-2017 dataset, measured in seconds. Considering the importance of achieving optimal results while also minimizing computational overhead, runtime performance becomes a crucial factor in model selection. Among the models exhibiting perfect F1 scores (i.e., RF, DT), Decision Tree (DT) emerges as the most time-efficient option, requiring approximately four minutes for training and testing combined. This makes it an attractive choice for scenarios where computational resources are limited. 

Moving beyond models with perfect F1 scores, Logistic Regression (LR) stands out as the fastest option among those achieving relatively good performance across all metrics. Conversely, models like Bagging (Bag) and Blending (Bled) demonstrate significantly longer runtimes, exceeding two hours. While these models may offer competitive performance, their computational demands make them less practical for resource-constrained applications. We also emphasize that when an IDS is online (in real time), the ensemble models are already trained, thus they can still be used for prediction. 

Additionally, it is worth noting that the dataset itself plays a significant role in determining runtime complexity. The CICIDS-2017 dataset is large and complex, which further accentuates the importance of efficient model selection. By analyzing the runtime performance of various models, stakeholders gain valuable insights into the computational demands associated with each algorithm, enabling informed decisions regarding model deployment and scalability.

\begin{table}[t!]
\centering
\vspace{1mm}
\caption{Performance of different models (individual ML models, and different ensemble methods) organized by F1 score (highest to lowest) on CICIDS-2017 Dataset.}
\resizebox{\linewidth}{!}{
\begin{tabular}{|l|c|c|c|c|}
\hline
\textbf{Models} & \textbf{Accuracy (ACC)} & \textbf{Precision (PRE)} & \textbf{Recall (REC)} & \textbf{F1 Score} \\
\hline
Random Forest (RF) & 1.00 & 1.00 & 1.00 & 1.00 \\
Bagging (Bag) & 0.998 & 1.00 & 1.00 & 1.00 \\
Blending (Bled) & 0.998 & 1.00 & 1.00 & 1.00 \\
Weighted Average (weighed\_avg) & 0.998 & 1.00 & 1.00 & 1.00 \\
Stacking & 0.997 & 1.00 & 1.00 & 1.00 \\
Gradient Boosting (GB) & 0.988 & 0.99 & 0.99 & 0.99 \\
Decision Tree (DT) & 0.998 & 0.998 & 0.998 & 0.998 \\
Cat Boost (CAT) & 0.998 & 0.998 & 0.998 & 0.998 \\
Multi-Layer Perceptron (MLP) & 0.996 & 0.996 & 0.996 & 0.996 \\
XGBoost (XGB) & 0.996 & 0.996 & 0.996 & 0.996 \\
Average (avg) & 0.996 & 0.997 & 0.996 & 0.996 \\
Max Voting (Max\_Vot) & 0.926 & 0.89 & 0.93 & 0.90 \\
Logistic Regression (LR) & 0.889 & 0.866 & 0.899 & 0.877 \\
AdaBoost (ADA) & 0.891 & 0.81 & 0.89 & 0.85 \\
\hline
\end{tabular}
}
\label{table:cic2017_all_features}
\end{table}

\begin{table}[t!]
\centering
\caption{Model Training and Testing Timetable (Seconds) for CICIDS-2017 Dataset. Logistic Regression and Decision Tree are the most efficient individual model while Max Voting is the most time-efficient ensemble learning method.}
\resizebox{0.9\linewidth}{!}{
\begin{tabular}{|l|c|}
\hline
\textbf{Models} & \textbf{Time (Seconds)} \\
\hline
Logistic Regression (LR) & 120.31 \\
Decision Tree (DT) & 240.7 \\
Max Voting (Max\_Vot) & 670.25 \\
Random Forest (RF) & 1402.8 \\
AdaBoost (ADA) & 2040.54 \\
Average (avg) & 2160.11 \\
XGBoost (XGB) & 2220.00 \\
Bagging (Bag) & 2305.31 \\
Cat Boost (CAT) & 2640.02 \\
Gradient Boosting (GB) & 3000.02 \\
Multi-Layer Perceptron (MLP) & 4200.45 \\
Weighted Average (weighed\_avg) & 5040.3 \\
Blending (Bled) & 9600.52 \\
Stacking & 9720.54 \\
\hline
\end{tabular}
}
\label{table:cic2017_time_table}
\end{table}


\subsubsection{Confusion Matrices for Different Individual and Ensemble Models for CICIDS-2017 Dataset}

We next show confusion matrices for the  14 different learners tested to classify various attacks in the CICIDS-2017 Dataset. Due to having higher number of attack classes here compared to RoEduNet-SIMARGL2021, confusion matrices are different for most models. We now provide them for different models along with their main insights.

\begin{figure}
    \centering
    \includegraphics[width= \linewidth]{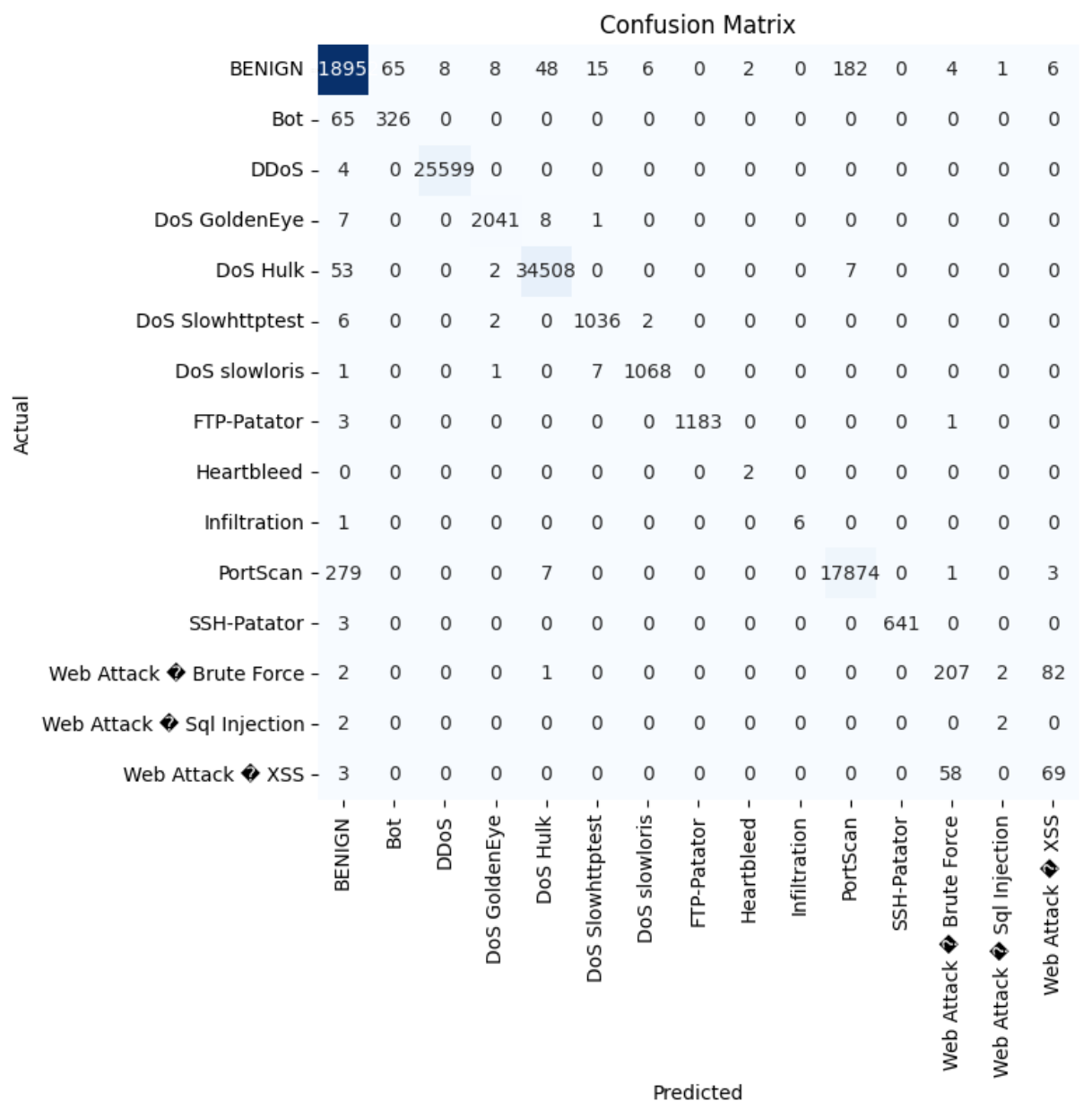}
    \caption{Confusion Matrix of Decision Tree (DT) on CICIDS-2017 Dataset. Most attacks are predicted very efficiently (except Web Attack and Bot).}
    \label{fig:confusion-matrix-DT-cic}
    \vspace{-4mm}
\end{figure}

\begin{figure}
    \centering
    \vspace{-9mm}\includegraphics[width=\linewidth]{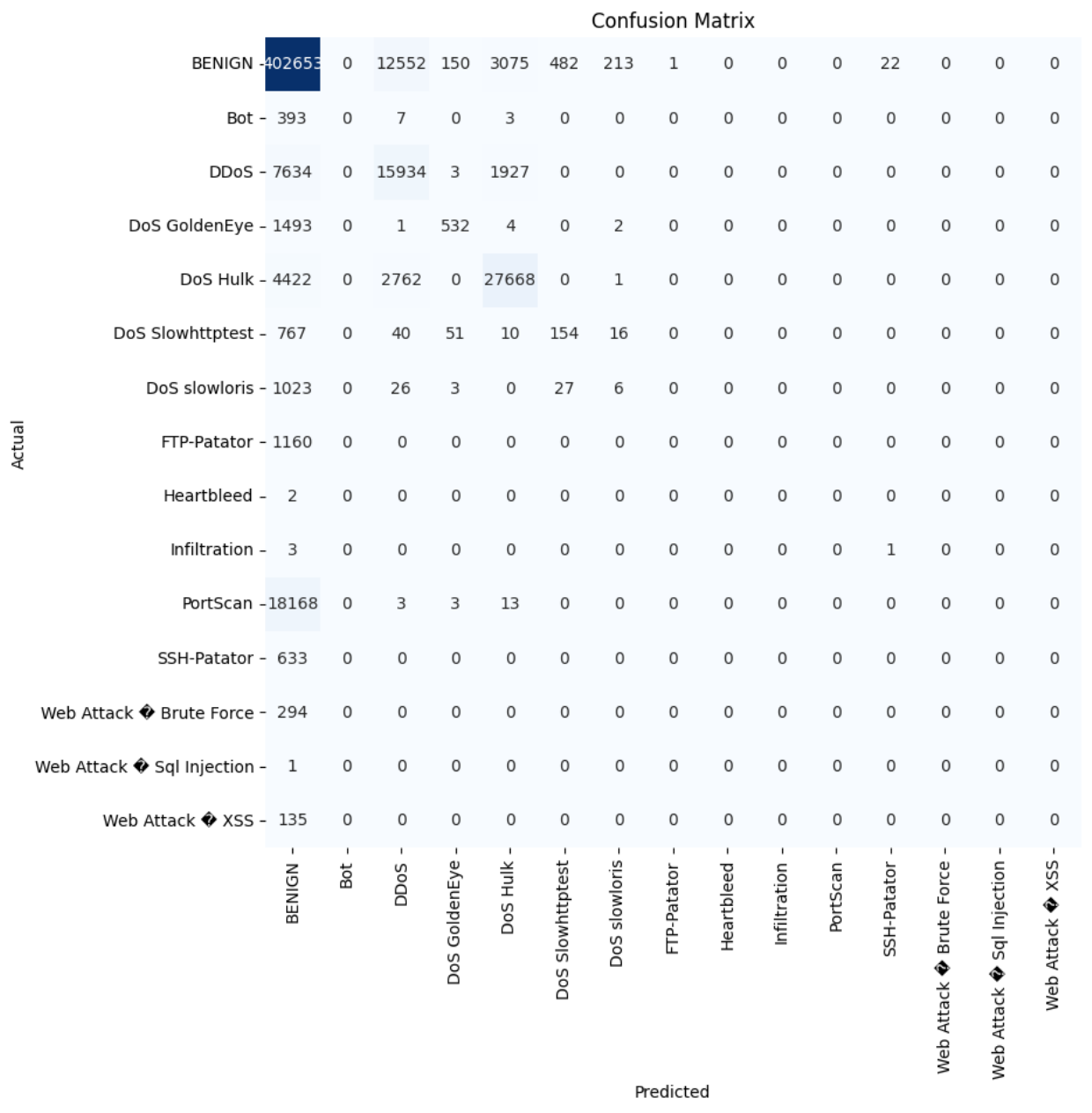}
    \caption{Confusion Matrix of Logistic Regression (LR) on CICIDS-2017 Dataset. It shows much lower prediction accuracy for different attacks.}
    \label{fig:confusion-matrix-lr-cic}
\end{figure}

\textbf{(i) Confusion Matrices of Individual Models on CICIDS-2017:}  Figure~\ref{fig:confusion-matrix-DT-cic} shows the confusion matrix of Decision Tree (DT) on CICIDS-2017 Dataset. It shows that most attacks are predicted very efficiently (except Web Attack and Bot). On the other hand, Figure~\ref{fig:confusion-matrix-lr-cic} shows the confusion matrix of Logistic Regression (LR) on CICIDS-2017, which shows much lower prediction accuracy.

\begin{figure}
    \centering    \vspace{-10mm}
\includegraphics[width=\linewidth]{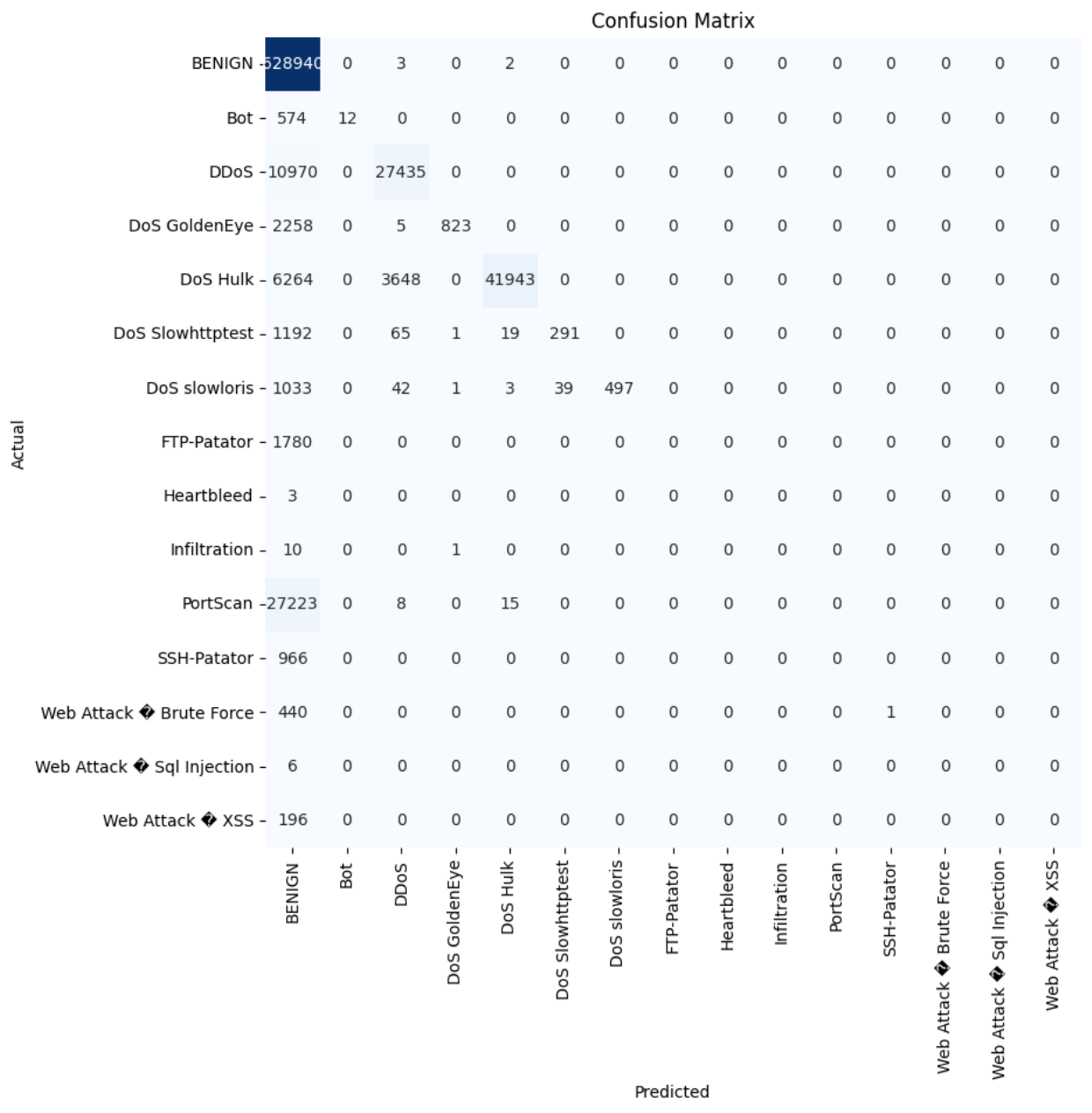}
    \caption{Confusion Matrix of Max Voting  (Max\_Vot) simple ensemble technique on CICIDS-2017 Dataset. It has lower prediction accuracy for several attacks.}
    \label{fig:confusion-matrix-Max_vot-cic}
\end{figure}

\begin{figure}
    \centering
    \vspace{-1mm}
\includegraphics[width=\linewidth]{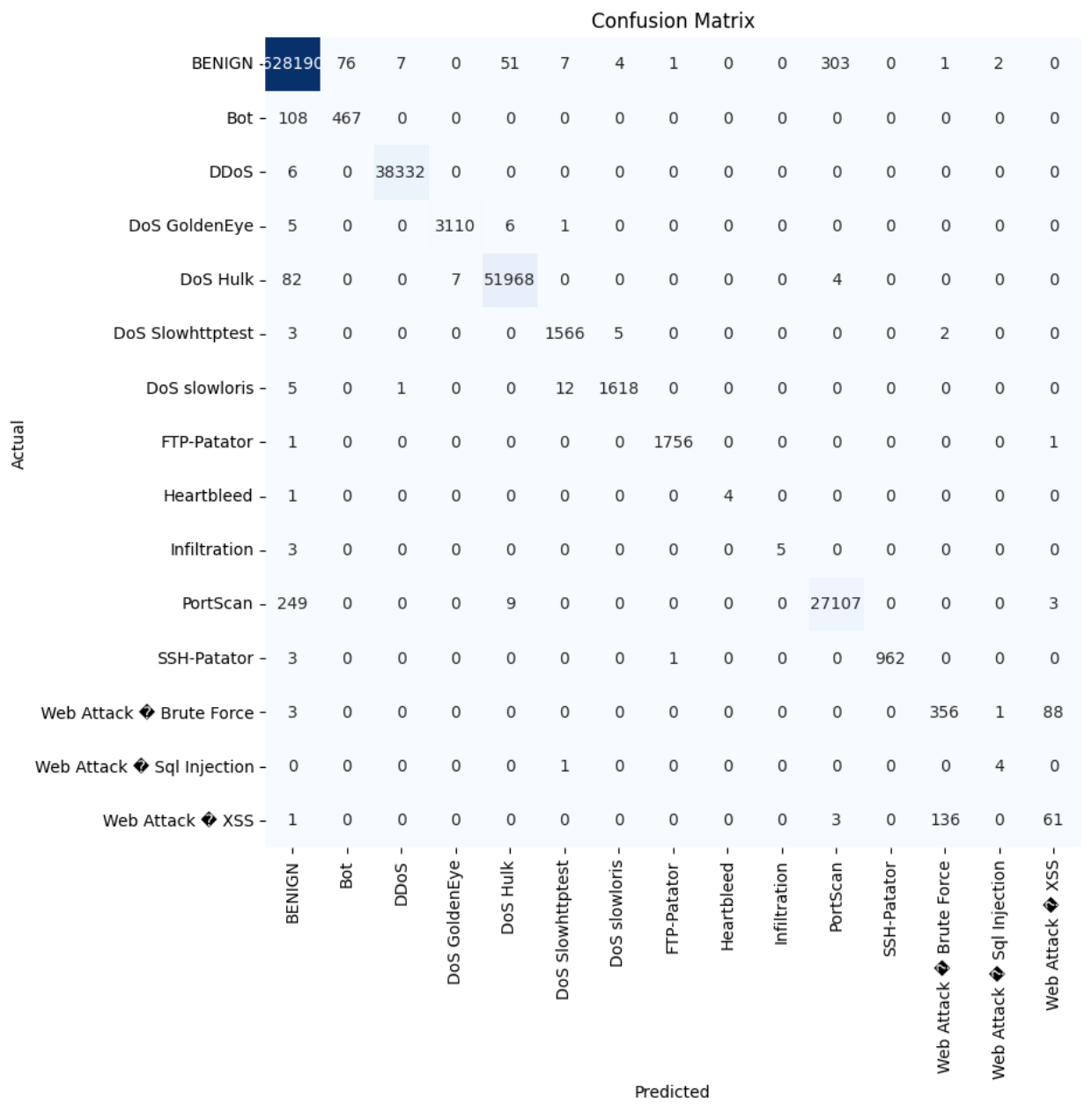}
    \caption{Confusion Matrix of Weighted Averaging  (Weighted\_avg) ensemble technique on CICIDS-2017 Dataset. Most attacks are predicted very efficiently (except Web Attack
and Bot).}
    \label{fig:confusion-matrix-Weighted_avg-cic}
    \vspace{-3mm}
\end{figure}

\textbf{(ii) Confusion Matrices of Simple Ensemble Methods on CICIDS-2017:} Figure~\ref{fig:confusion-matrix-Max_vot-cic} shows the confusion
matrix of Max Voting (Max\_Vot) on CICIDS-2017, which shows lower prediction accuracy for several attacks. On the contrary, Figure~\ref{fig:confusion-matrix-Weighted_avg-cic} shows the confusion matrix of
Weighted Averaging (Weighted\_avg) on CICIDS-2017 Dataset. It shows that
most attacks are predicted very efficiently (except Web Attack
and Bot). The Averaging simple ensemble method has also same behaviour as Weighted Averaging (figure omitted).

\begin{figure}
    \centering
    \includegraphics[width=\linewidth]{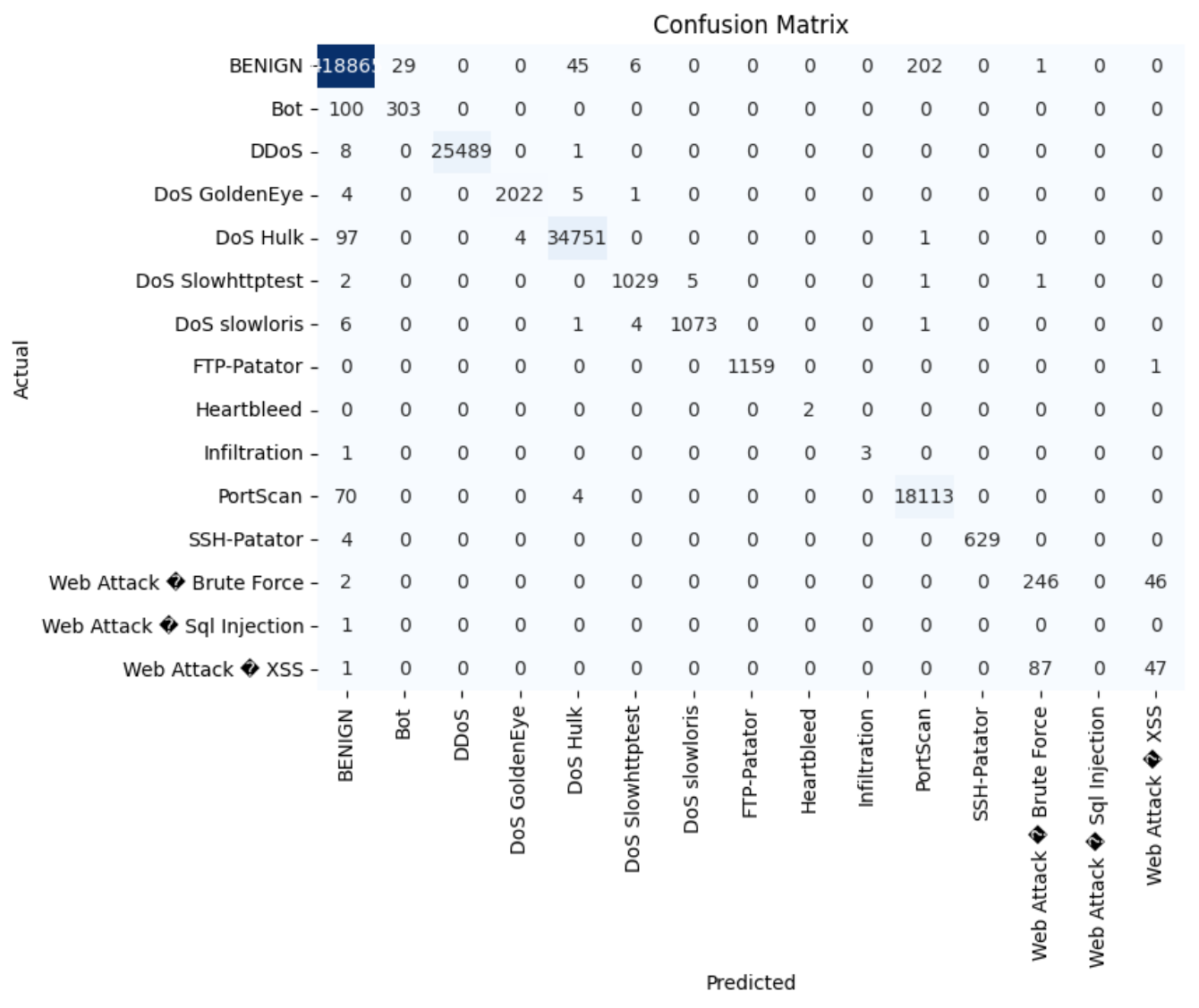}
    \caption{Confusion Matrix of Bagging (Bag) ensemble technique on CICIDS-2017 Dataset. Most attacks are predicted very efficiently (except
Web Attack-XSS and Bot).}
    \label{fig:confusion-matrix-Bag-cic}
\end{figure}

\begin{figure}
    \centering
    \includegraphics[width=\linewidth]{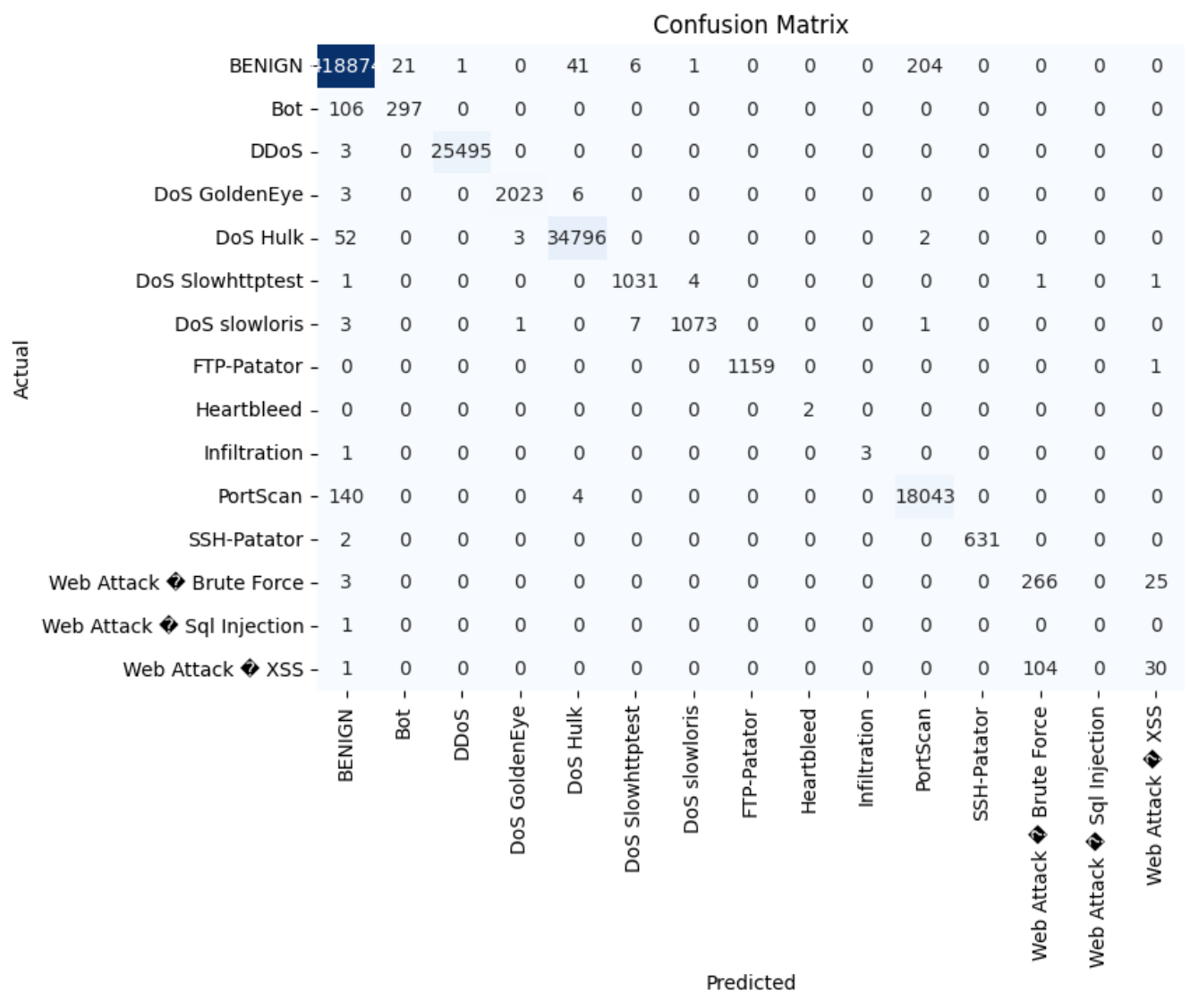}
    \caption{Confusion Matrix of Blending  (Bled) ensemble technique on CICIDS-2017 Dataset. Most attacks are predicted very efficiently (except
Web Attack-XSS and Bot).}
    \label{fig:confusion-matrix-Bled-cic}
\end{figure}

\begin{figure}
    \centering
    \includegraphics[width=\linewidth]{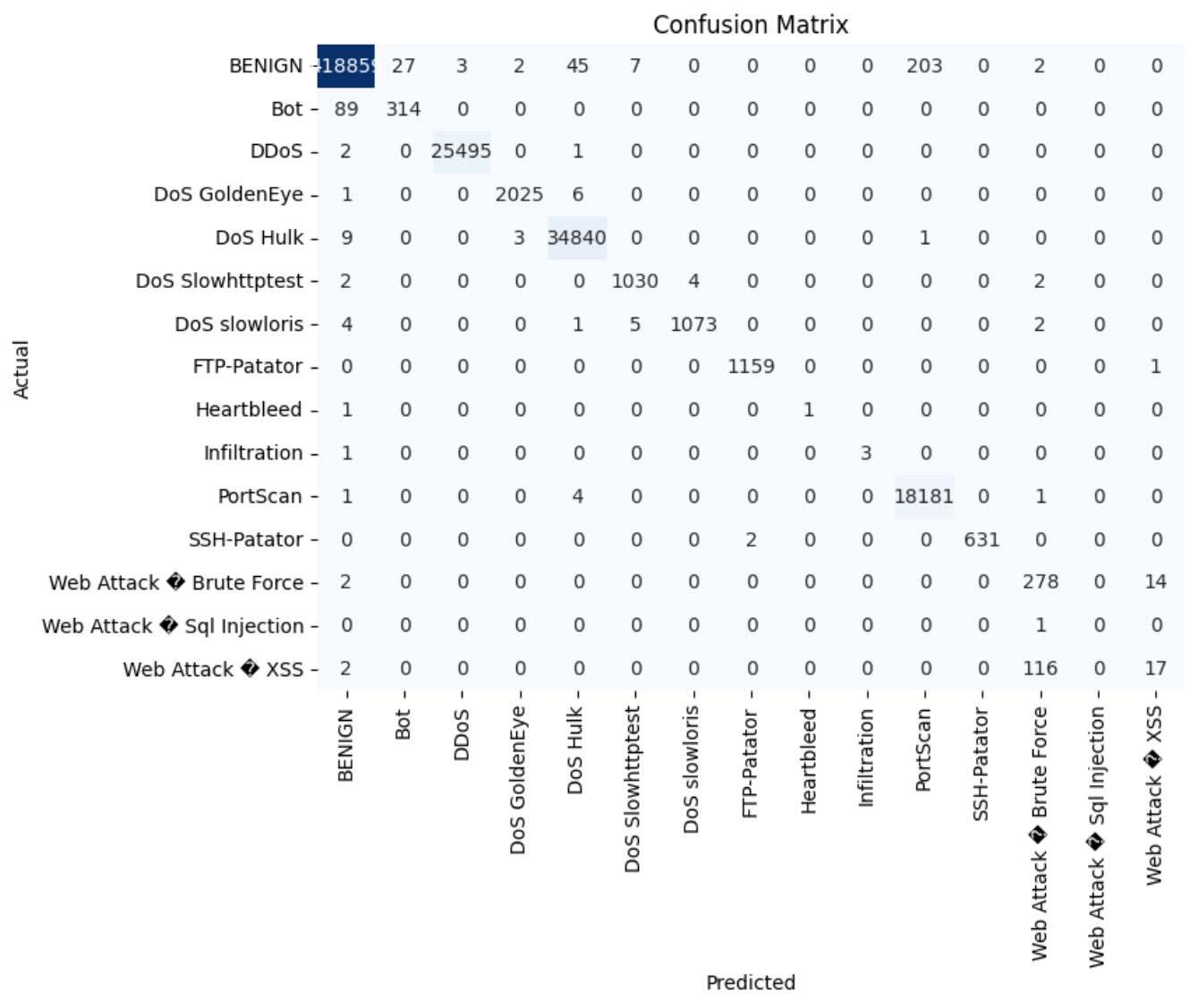}
    \caption{Confusion Matrix of Cat Boosting (CAT) ensemble technique on CICIDS-2017 Dataset. It has near-perfect performance for all intrusion classes except Bot.}
    \label{fig:confusion-matrix-CAT-cic}
\end{figure}

\begin{figure}
    \centering    \includegraphics[width=\linewidth]{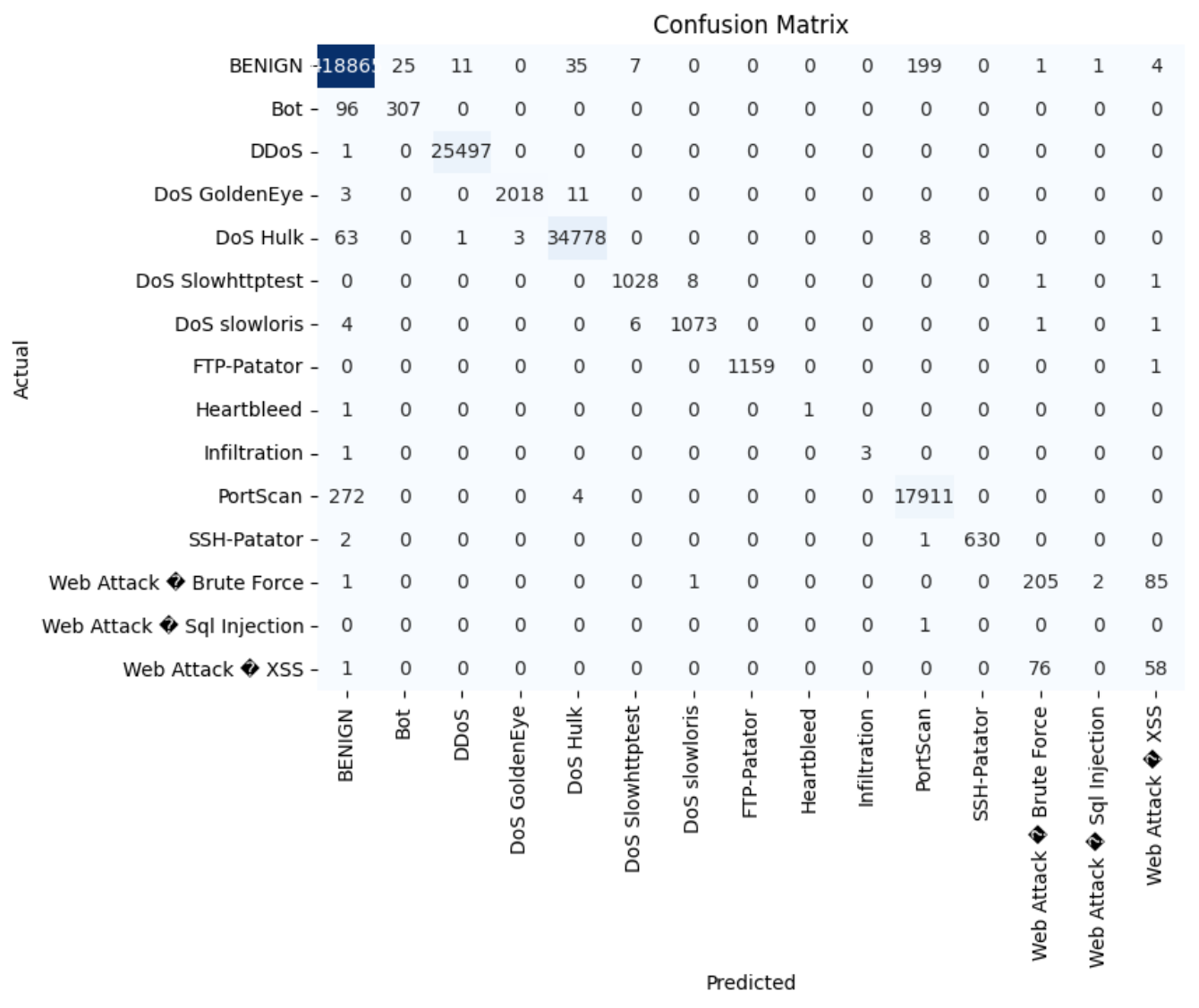}
    \caption{Confusion Matrix of Stacking ensemble technique on CICIDS-2017 Dataset. Most attacks are predicted very efficiently (except Web Attack-XSS and Bot).}
    \label{fig:confusion-matrix-Stack-cic}
\end{figure}

\begin{figure}
    \centering
    \includegraphics[width=\linewidth]{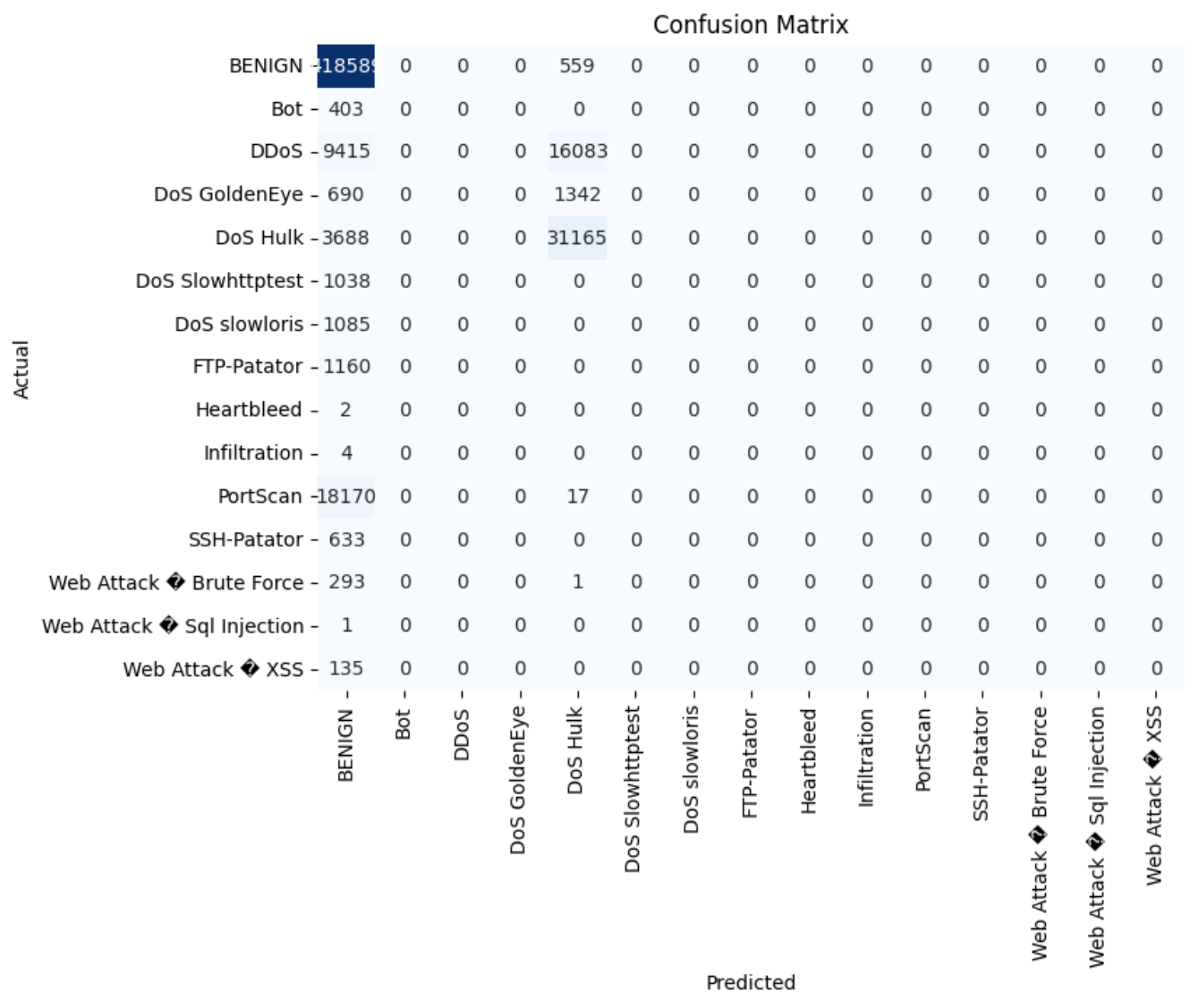}
    \caption{Confusion Matrix of Adaptive Boosting (ADA) ensemble technique on CICIDS-2017 Dataset. It has the lowest performance across all models.}
    \label{fig:confusion-matrix-ADA-cic}
    \vspace{-3mm}
\end{figure}

\begin{figure}
    \centering
    \includegraphics[width=\linewidth]{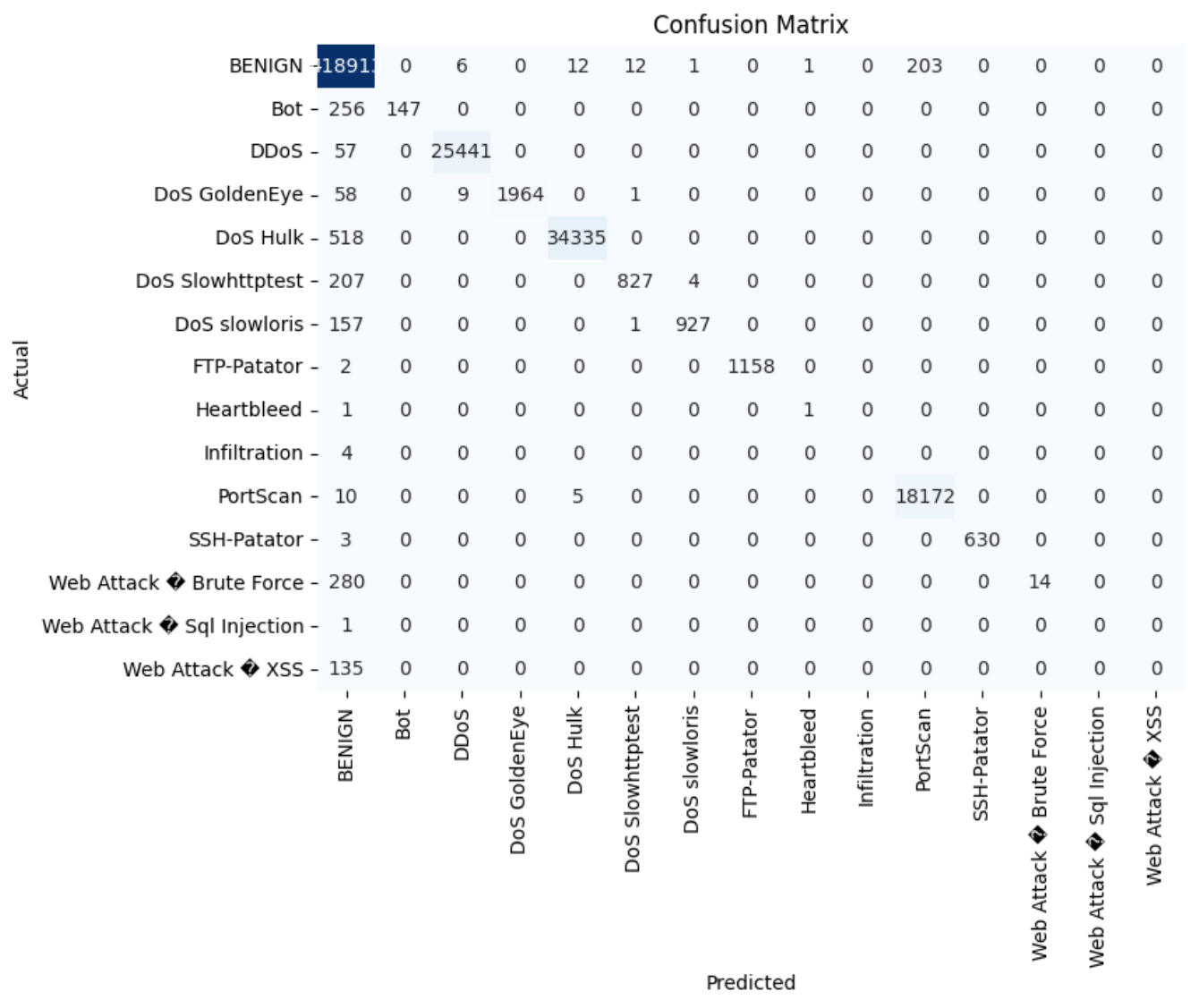}
    \caption{Confusion Matrix of Extreme Gradient Boosting (XGB) ensemble technique on CICIDS-2017 Dataset. It has lower performance compared to other advanced ensemble methods (except ADA).}
    \label{fig:confusion-matrix-XGB-cic}
\end{figure}

\textbf{(iii) Confusion Matrices of Advanced Ensemble Methods on CICIDS-2017 Dataset:} Figures~\ref{fig:confusion-matrix-Bag-cic}-\ref{fig:confusion-matrix-Bled-cic} show the confusion matrices of Bagging (Bag) and Blending (Bled) on CICIDS-2017 Dataset, respectively. They show that both ensemble methods predict
most attacks very efficiently (except Web Attack-XSS and Bot). Along the same lines, Figure~\ref{fig:confusion-matrix-CAT-cic} shows the confusion matrix for Cat Boosting (CAT) ensemble technique on CICIDS-2017 Dataset. It has near-perfect performance for all intrusion classes except Bot. Similarly, Figure~\ref{fig:confusion-matrix-Stack-cic} shows that Stacking ensemble method predict most attacks very efficiently (except Web Attack-XSS and Bot). On the other hand, Boosting ensemble learning techniques provide lower prediction accuracy for different classes for the CICIDS-2017 dataset (as shown in Figures~\ref{fig:confusion-matrix-ADA-cic}-\ref{fig:confusion-matrix-XGB-cic}). However, we emphasize that Extreme Gradient Boosting Ensemble Techniques (XGB) has better performance compared to Adaptive Boosting (ADA).

\subsection{Performance Enhancement Across Datasets }

\textbf{CICIDS-2017:} Our analysis on the CICIDS-2017 dataset revealed significant gains achieved through the implementation of our framework. The RF was the best model that demonstrated a notable improvement in accuracy, precision, recall (achieving a perfect score of 1.000), and F1 score. Furthermore, leveraging this base learner in conjunction with the best ensemble learning methods (including bagging, blending, weighted averaging, and stacking) yield near-perfect performance metrics, including accuracy and precision, while maintaining a perfect recall and F1 score. 

\textbf{RoEduNet-SIMARGL2021:} Similarly, on the RoEduNet-SIMARGL2021 dataset, employing our framework yielded remarkable improvements. The best ensemble learning model is Averaging that exhibited achieving perfect scores of 1.000 across all performance metrics, including accuracy, precision, recall, and F1 score. Similarly, Max Voting, Stacking, Bagging, Boosting, and Weighted Averaging yield almost perfect performance across all four metrics. 

Overall, our comprehensive evaluation showcases the potential benefits of our proposed simple and advanced ensemble learning. The detailed breakdown of performance metrics for each dataset underscores the efficacy of our approach and its ability to enhance classification accuracy and reliability across diverse datasets.

We also emphasize that one strong point in this work is categorizing the different individual models and ensemble methods based on their confusion matrices (which show different aspects of model's performance on different intrusion classes). In particular, these confusion matrices can give insights about which model can be used depending on the expected types of network attacks for the organization that the security analyst is monitoring. This also can lead to doing ``ensemble of ensembles,'' where we fuse different ensemble methods to confront different network attacks (where each ensemble method has some strong performance on one or more of these attacks).

Having finished our detailed evaluation analysis on our two datasets, we next show the main discussions and limitations of our work in the next section.

%% file: Discussion.tex
\section{Limitations, Discussion, and Future Directions}\label{sec: Discussion}
 
\subsection{DISCUSSION}
\subsubsection{Significance of our Framework} 
In the contemporary landscape of flourishing information, the frequency of network attacks is expected to rise (as evidenced by recent studies such as the one conducted by the Center for Strategic \& International Studies (CSIS)~\cite{cyber-attacks-decade}).
Despite the evolution of Intrusion Detection Systems (IDS), security analysts still face the challenge of verifying potential attack incidents in this rapidly evolving environment. Hence, having a reliable framework for intrusion detection systems can significantly mitigate this challenge by reducing the instances of false positive rates (FPR) to scrutinize and enable a focused analysis of critical traffic data. The framework presented in our study contributes to addressing this issue via enhancing the performance metrics (including accuracy, recall, precision, and F1 score) of intrusion detection systems, which is pivotal for their deployment to ensure better  network security in aforementioned modern systems. Our framework was also evaluated using  confusion matrices (which show intrusion-specific performance of different
individual methods and ensemble approaches). The time analysis also can help in choosing the best model, depending on the needs of the security analyst. 

\subsubsection{Summary of Results} 
All findings presented in this paper are succinctly outlined in Section~\ref{sec:evaluation}, addressing key inquiries readers may have, such as determining the optimal AI model and ensemble technique and assessing their viability. 
Section~\ref{sec:evaluation} provides a comprehensive overview of the results, highlighting performance metrics and crucial aspects of the ensemble technique, including runtime and confusion metrics, employed in evaluating the framework.

\textbf{Performance Metrics:} This section focuses on metrics such as Accuracy, Precision, Recall, and F1 scores utilized to assess individual models and ensemble techniques. Notably, superior results, indicated by higher metric scores, were observed for models and ensemble techniques including Random Forest (RF), Decision Tree (DT), Weighted Average (weighed\_avg), Stacking, Bagging (Bag), Blending (Bled), AdaBoost (ADA), Cat Boost (CAT), Gradient Boosting (GB), and XGBoost (XGB).

\textbf{Runtime:} The runtime analysis involves evaluating the execution times of the 14 individual models and ensemble techniques utilized. In Section~\ref{sec:evaluation}, a runtime table is provided, arranged from the fastest to the slowest models. Following this analysis, the fastest models overall include LR, DT, MLP, and RF. Models with an average runtime encompass Bag, XGB, ADA, CAT, and GB. Conversely, the slowest models comprise Bled, avg, stacking, max\_Vot, and weighted\_avg.

\textbf{Optimal Models:} By intersecting these results (performance metrics and runtime), we identify the optimal models, including DT, RF, Bag, Bled, ADA, CAT, and GB. Subsequently, Stacking, avg, weighted\_avg, and Max\_Vot are also recognized, albeit with slower runtimes. These models demonstrated superior performance across the metrics outlined in this study.

\textbf{Superiority of Ensemble Learning Methods:} Table~\ref{table:cic2017_top_3_ensemble_models} presents a  summary of top-5 models (in terms of F-1 score) for CICIDS-2017
and RoEduNet-SIMARGL2021 Datasets. We notice that the ensemble methods have superiority over individual models for both datasets. In particular, for CICIDS-2017 dataset, all the top-5 models are ensemble methods. On the other hand, for RoEduNet-SIMARGL2021 dataset we have four ensemble methods in the top-5 methods (where decision tree (DT) is the only individual model in this list of top-5 models).


\begin{table}[t!]
\centering
\vspace{1mm}
\caption{A summary of top-5  models for CICIDS-2017 and RoEduNet-SIMARGL2021 datasets.}
\resizebox{\linewidth}{!}{
\begin{tabular}{|l|c|c|}
\hline
\textbf{Ranking} & 
\textbf{CICIDS-2017} & \textbf{RoEduNet-SIMARGL2021} \\
\hline
1st & Random Forest (RF) & Random Forest (RF) \\
2nd & Bagging (Bag) & Decision Tree (DT) \\
3rd & Blending (Bled) & Averaging (avg) \\
4th & Weighted Averaging  & Max Voting \\
5th & Stacking  & Stacking \\
\hline
\end{tabular}
}
\label{table:cic2017_top_3_ensemble_models}
\end{table}

\subsubsection{Random Forest Assessment} The findings highlight Random Forest (RF) as one of the top-performing ensemble techniques across both datasets. However, its applicability may be limited by a tendency towards overfitting and bias when deployed in diverse scenarios. For instance, concerning the CICIDS-2017 dataset, the most effective base learner is decision tree (DT), which achieved an Accuracy of 0.998, Precision of 1.0, Recall of 1.0, and F1 score of 1.0. Similarly, for the RoEduNet-SIMARGL2021 dataset, the next best-performing base learner is decision tree (DT), attaining an Accuracy of 1.0, Precision of 1.0, Recall of 1.0, and F1 score of 1.0.

\subsubsection{Advantages of Ensemble Learning Framework} 

The ensemble learning framework offers versatility in its construction approach.  Our setup facilitates a comprehensive evaluation of the two datasets utilized, including RoEduNet-SIMARGL2021, which, to the best of our knowledge, has been evaluated in this context with very limited works within the existing literature~\cite{10540382}. However, our current work
has several differences from the prior work~\cite{10540382}. First, our work considers all the individual classes of CICIDS-2017 (total of 15 classes), and RoEduNet-SIMARGL2021 (total of 4 classes). The previous work simplified the problem for CICIDS-2017 grouping the attacks in 7 classes, and by not using the malware class for RoEduNet-SIMARGL2021. Moreover, this current work presents in detail the decision matrices used for each case scenario where it brings light to the classes that ensemble learning can identify the best, and other attacks that might need enhancement to detect. This important analysis is not present in the previous work~\cite{10540382} since it considered the problem from a holistic point of view. Furthermore, this current work employs the blending technique which is not present in the prior work~\cite{10540382}. Finally, the work~\cite{10540382} focused on hierarchical ensemble learning where it built a two-level ensemble learning framework for network intrusion detection tasks, which is different from our current focus of comprehensive comparison of popular ensemble methods.

Furthermore, this study presents an extensive evaluation that distinguishes it from previous works by analyzing 14 models/ensemble methods across two distinct datasets, yielding results for Accuracy, Precision, Recall, F1, Confusion Matrices, and Runtime. Notably, we achieve near-perfect results for several models in terms of F1 score, emphasizing the significance of these metrics for Intrusion Detection Systems (IDS). Given the imperative for security analysts, stakeholders, and users to accurately and rapidly identify potential threats, undetected attacks which  can pose substantial risks.

It is worth emphasizing that we have taken the extra step of making our codes open source. Designed to be easily adaptable for use with other datasets and further analysis, they do not constitute a deployable solution for production, as they have not undergone extensive testing or validation by an authoritative entity. Instead, they serve as a proof of concept highlighting the benefits of our proposed framework and represent a crucial step towards enhancing the field of AI-based network IDS. 

\subsection{LIMITATIONS}

\subsubsection{Dataset Analysis and Biases} 
The experiments conducted enable us to draw several noteworthy conclusions regarding the datasets employed. Firstly, it is evident that the models within our framework achieved superior gains when applied to CICIDS-2017 and RoEduNet-SIMARGL2021 datasets. Each dataset exhibits distinct characteristics. For example, RoEduNet-SIMARGL2021 comprises nearly 30 million data points with approximately 20 feature columns, whereas CICIDS-2017 contains almost 2 million data points but approximately 70 feature columns (refer to Table~\ref{tbl:combined_dataset_statistics}). This discrepancy accounts for the observed increase in runtime for these datasets. Additionally, the disparity in data volume and class count is noteworthy, with RoEduNet-SIMARGL2021 featuring four prediction classes and CICIDS-2017 incorporating several prediction classes. Interestingly, AI models appear to readily learn patterns in the RoEduNet-SIMARGL2021 dataset owing to its extensive size and fewer prediction classes. Conversely, despite the higher number of prediction classes in CICIDS-2017, models demonstrate adeptness and achieve commendable scores, possibly due to its heavily unbalanced class distribution (as depicted in Table~\ref{tbl:combined_dataset_statistics}, where four classes combined represent less than 1\% of the entire dataset). We stress here that this is close to a real-world scenario since most of the traffic is normal traffic. This limitation underscores the necessity for future research to explore alternative datasets or employ uncalibrated models  to use on ensemble learning to broaden benchmarking and testing within our framework.


\subsection{FUTURE DIRECTIONS}

While our work represents a foundational step towards advancing AI-based Intrusion Detection Systems (IDS), there are numerous avenues for future exploration and refinement within our framework. Expanding our framework to encompass additional datasets, diverse AI models, and a broader array of ensemble methods holds promise for creating a more comprehensive and insightful framework. 

Another promising avenue for future research involves delving into multi-level ensemble learning approaches, which involve utilizing multi-level classification techniques to further enhance detection accuracy and robustness. Additionally, exploring a wider range of feature selection methods could provide valuable insights into optimizing model performance and interpret ability.

Moreover, integrating explainable AI (XAI) frameworks to generate explanations for ensemble methods presents an intriguing direction for future investigation. By providing transparent and interpretable explanations for model predictions, XAI techniques could enhance the trust and understanding of IDS systems~\cite{mahbooba2021explainable}.

Furthermore, the ultimate goal of our ongoing efforts includes the implementation of real-time capabilities and validation through collaboration with security experts and analysts. This collaboration aims to gather invaluable insights and feedback, leading to continuous improvements and real-world applicability of our framework.

%% file: Conclusion.tex
\section{Conclusion}\label{sec: conclusion}

The primary goal of a security intrusion detection tool is to serve as a robust shield against potential intruders. Leveraging artificial intelligence (AI) can significantly enhance the automation and effectiveness of these tools. The increasing frequency of intrusions in networked systems has driven extensive research into developing AI techniques for intrusion detection systems (IDS). While various AI models have been deployed for this purpose, each model has its own strengths and weaknesses, presenting a challenge in selecting the most suitable model for a given dataset.

To address this challenge, combining multiple AI models can substantially improve their overall performance and applicability in network intrusion detection. This paper aims to bridge this crucial gap by evaluating a diverse array of ensemble methods for IDS. Specifically, we present a comprehensive comparative study of individual models and both simple and advanced ensemble learning frameworks for network intrusion detection tasks. Our approach involves training base learners and ensemble methods to generate evaluation metrics.

We present results for 14 combinations of individual and ensemble models within our framework, utilizing various boosting, stacking, and blending methods on diverse base learners. Evaluation is conducted on two network intrusion datasets, each possessing unique characteristics. Our analysis categorizes AI models based on their performance metrics (including accuracy, precision, recall, and F1-score), and runtime, highlighting the advantages of learning across various setups for two very important datasets.


The best models for each dataset were advised based on their performance metrics. For the CICIDS-2017 dataset, the top three ensemble models were Random Forest (RF), Bagging (Bag), and Blending (Bled). These models achieved exceptional results, with the Random Forest model achieving perfect scores in Accuracy (ACC), Precision (PRE), Recall (REC), and F1 Score. Bagging and Blending models also performed remarkably well, achieving near-perfect metrics across the board (see Table \ref{table:cic2017_top_3_ensemble_models}). Similarly, for the RoEduNet-SIMARGL2021 dataset, the top three models were Random Forest (RF), Decision Tree (DT), and Bagging (Bag). Both the Random Forest and Decision Tree models achieved perfect scores in all performance metrics, while the Bagging model performed almost perfectly (see Table \ref{table:cic2017_top_3_ensemble_models}).

Our evaluation results show that using ensemble learning was beneficial as it significantly enhanced the performance of the models, leading to high accuracy, precision, recall, and F1 scores across both datasets.

We contribute to the community by providing our source codes, offering a foundational ensemble learning framework for network intrusion detection that can be expanded with new models and datasets. We also provide insights into the best models for each dataset, highlighting common and distinct behaviors among them through confusion matrices, which influence their performance and results. We conclude with an in-depth discussion of our main findings and the primary benefits of our framework. This study represents a significant advancement in utilizing ensemble learning methods for network Intrusion Detection Systems (IDS), achieved through comprehensive evaluations and comparisons of various metrics to assess the effectiveness of these ensemble methods.

%% file: Appendix.tex
\section{AI Models and Hyper-parameters}\label{app:ai_models_hyperparams}

We present the hyperparameters of the various AI models and ensemble methods employed in this study.

\subsection{Details of AI Models and Hyperparameters}

\subsubsection{Base Models}
First, we outline the primary details of the base models.

\textbf{Logistic Regression (LR):} Moving on to the next classifier, we employed Logistic Regression. The parameter configuration for this classifier remains at default settings.

\textbf{Decision Tree (DT):} Continuing with the subsequent classifier, we utilized the Decision Tree. The parameter configuration for this classifier remains at default settings.


\textbf{Multi-layer Perceptron (MLP):} Following, we utilized the MLP classifier with the subsequent configuration:
The MLP classifier architecture comprises two hidden layers, each containing 50 neurons, utilizing the Rectified Linear Unit (ReLU) activation function. We employed the Adaptive Moment Estimation (Adam) solver for optimization, with an L2 regularization term (alpha) set to 0.0001. The batch size was dynamically adjusted based on the dataset size. Additionally, the learning rate was kept constant at 0.001 throughout training, with a maximum of 1000 iterations. The random seed was fixed at 42 for reproducibility. Early stopping was disabled, and progress messages were printed during training.\vspace{-2mm}

\subsubsection{Ensemble Methods}

Next, we present the key details of our ensemble methods.

\textbf{AdaBoost (ADA)}: AdaBoost was employed as the next classifier. The parameter configuration for this classifier remains at the default setting.

\textbf{Extreme Gradient Boost (XGB)}: Following, XGB was utilized as a classifier. The parameter configuration learning rate set to 0.1, loss function set to multi: softmax.

\textbf{CatBoost (CAT)}: Subsequently, Catboost was utilized as a classifier. The parameter settings default setting.

\textbf{Max Voting}: The following classifier employed is Voting, a simple stacking method that aggregates each model's decision. In this implementation, a VotingClassifier is instantiated with two base classifiers, Logistic Regression (LR) and Decision Tree (dt), using hard voting. 


\textbf{Average}: Additionally, the Average classifier was employed. This approach involves initializing three base classifiers: Decision Tree, K-nearest Neighbors, and Random Forest. These models are trained on the training data. The predictions from each model (pred1, pred2, pred3) are then averaged to generate the final prediction using a simple averaging technique.

\textbf{Weighted Average}: Weighted Average was utilized in this implementation. This method involves initializing three base classifiers: Decision Tree, K-nearest Neighbors, and Random Forest. An ensemble is then created using a VotingClassifier, with the classifiers assigned weights based on their importance. In this case, Decision Tree is assigned a weight of 0.4, K-nearest Neighbors a weight of 0.3, and Random Forest a weight of 0.3.

\textbf{Bagging}: Bagging, a class of ensemble methods, was utilized in this scenario. This method involves dividing the dataset into subsets with replacements and utilizing these subsets as input data for diverse base models. Subsequently, the predictions from each base model are aggregated to reach a final decision. In this implementation, Bagging Classifier was instantiated with various base models including Random Forest, MLP Classifier, Logistic Regression, and Decision Tree Classifier. The number of estimators for the Bagging Classifier was set to the total number of base models, ensuring that each base model contributes to the ensemble's prediction.

\textbf{Random Forest (RF):} One ensemble classifier utilized for detecting malicious samples in network traffic was the Random Forest (RF). The hyperparameters employed for this classifier are as follows: n\_estimators (representing the number of trees used) was set to 100, the maximum tree depth was set to 10, the minimum number of samples required to split an internal node was set to 2, while the remaining parameters were left as default. 



\textbf{Blending}: Blending, another class of ensemble methods, was employed. This method uses a holdout (validation) set from the train set to make predictions. The process involves splitting the train set into training and validation sets, fitting models on the training set, and making predictions on the validation and test sets. The validation set and its predictions are then used to build a new model, which makes the final decision on the test set. In this implementation, the blending method was applied using several base models, including Random Forest, Multi-layer Perceptron, Logistic Regression, and Decision Tree. The predictions from these base models on the validation set are used to train the final estimator, which makes the final decision.

\textbf{Stacking}: Lastly, Stacking, another class of ensemble methods, was employed. This method stacks the decisions of base models, utilizing their outcomes to create a new dataset. Subsequently, various models and ensemble methods are applied again to make a final decision. In this implementation, StackingClassifier was instantiated with several base models, including Random Forest, Multi-layer Perceptron, Logistic Regression, and Decision Tree. The predictions from these base models are then used to train the final estimator, which makes the final decision.